\documentclass{aa}
\usepackage{amsmath,amssymb}
\usepackage[overload]{textcase}
\usepackage{rotating}
\usepackage{afterpage}
\usepackage{tabularx}
\usepackage{hyperref}
\usepackage{multirow}
\usepackage{booktabs}
\usepackage{todonotes}
\usepackage{mathabx}

\def\arcsec{{^{\prime\prime}}}

\def\farcs{\hbox{$.\!\!^{\prime\prime}$}}

\def\gtrsim{\mathrel{\hbox{\rlap{\hbox{\lower4pt\hbox{$\sim$}}}\hbox{$>$}}}}
\def\lessim{\mathrel{\hbox{\rlap{\hbox{\lower4pt\hbox{$\sim$}}}\hbox{$<$}}}}

\begin{document}

   \title{Early Planet Formation in Embedded Disks (eDisk) XVI: An asymmetric dust disk driving a multi-component molecular outflow in the young Class 0 protostar GSS30 IRS3}
   \author{Alejandro Santamar\'ia-Miranda\inst{1}
          \and
          Itziar de Gregorio-Monsalvo\inst{1}
          \and
          Nagayoshi Ohashi\inst{2}  
          \and
          John J. Tobin\inst{3}
          \and
          Jinshi Sai \inst{2}
          \and
          Jes K. J\o rgensen\inst{4}
          \and
          Yusuke Aso\inst{5}
          \and
          Zhe-Yu Daniel Lin\inst{6}
          \and
          Christian Flores\inst{2}
          \and
          Miyu Kido\inst{7}
          \and
          Patrick M. Koch\inst{2}
          \and
          Woojin Kwon\inst{8,9}
          \and
          Chang Won Lee\inst{5,10}
          \and
          Zhi-Yun Li\inst{6}
          \and
          Leslie W. Looney\inst{3,11}
          \and
          Adele L. Plunkett\inst{3}
          \and
          Shigehisa Takakuwa\inst{2,7}
          \and
          Merel L.R  van 't Hoff\inst{12}
          \and
          Jonathan P. Williams\inst{13}
          \and
          Hsi-Wei Yen\inst{2}
        }

   \institute{European Southern Observatory, 3107, Alonso de C\'ordova, Santiago de Chile\\
              \email{alejandrosantamariamiranda@gmail.com}
         \and
        Academia Sinica Institute of Astronomy \& Astrophysics, 11F of Astronomy-Mathematics Building, AS/NTU, No.1, Sec. 4, Roosevelt Rd, Taipei 10617, Taiwan, R.O.C.
        \and
        National Radio Astronomy Observatory, 520 Edgemont Rd, Charlottesville, VA, 22903, USA
        \and
        Niels Bohr Institute, University of Copenhagen, {\O}ster Voldgade 5-7, 1350 Copenhagen K., Denmark
        \and
        Korea Astronomy and Space Science Institute, 776 Daedeok-daero, Yuseong-gu, Daejeon 34055, Republic of Korea
        \and
        University of Virginia, 530 McCormick Rd., Charlottesville, Virginia 22904, USA
        \and
        Department of Physics and Astronomy, Graduate School of Science and Engineering, Kagoshima University, 1-21-35 Korimoto, Kagoshima,Kagoshima 890-0065, Japan
        \and
        Department of Earth Science Education, Seoul National University, 1 Gwanak-ro, Gwanak-gu, Seoul 08826, Republic of Korea
        \and
        SNU Astronomy Research Center, Seoul National University, 1 Gwanak-ro, Gwanak-gu, Seoul 08826, Republic of Korea
        \and
        Division of Astronomy and Space Science, University of Science and Technology, 217 Gajeong-ro, Yuseong-gu, Daejeon 34113, Republic of Korea
        \and
        Department of Astronomy, University of Illinois, 1002 West Green St, Urbana, IL 61801, USA
        \and
        Department of Astronomy, University of Michigan, 1085 S. University Ave., Ann Arbor, MI 48109-1107, USA
        \and
        Institute for Astronomy, University of Hawai‘i at Mānoa, 2680 Woodlawn Dr., Honolulu, HI 96822, USA
        }

   \date{Received 14 March 2024; accepted 29 July 2024}

\abstract
{We present the results of the ALMA Large Program Early Planet Formation in Embedded disks observations of the Class 0 protostar GSS30 IRS3. Our observations included 1.3 mm continuum with a resolution of 0\farcs05 (7.8 au) and several molecular species including  $^{12}$CO, $^{13}$CO, C$^{18}$O, H$_{2}$CO and c-C$_{3}$H$_{2}$. The dust continuum analysis unveiled a disk-shaped structure with a major axis size of $\sim$200 au. We observed an asymmetry in the minor axis of the continuum emission suggesting that the emission is optically thick and the disk is flared. On the other hand, we identified two prominent bumps along the major axis located at distances of 26 and 50 au from the central protostar. The origin of the bumps remains uncertain and might be due to an embedded substructure within the disk or the result of the temperature distribution instead of surface density due to optically thick continuum emission. The $^{12}$CO emission reveals a molecular outflow consisting of three distinct components: a collimated one, an intermediate velocity component exhibiting an hourglass shape, and a wider angle low-velocity component. We associate these components with the coexistence of a jet and a disk-wind. The C$^{18}$O emission traces both a Keplerian rotating circumstellar disk and the infall of the rotating envelope. We measured a stellar dynamical mass of 0.35$\pm$0.09 M$_{\odot}$.} 

\maketitle

\section{Introduction}
\label{sec:p6_intro}
Circumstellar disks arise as a direct consequence of the angular momentum conservation of rotating material from an infalling envelope  \citep{Shu1984, Shu1987}. In these disks, planets are expected to form at some point during the star formation process. In the last decade, the efforts to understand planet formation have been focused mainly on Class II disks which have dispersed their envelopes. The advent of the Atacama Large Millimeter/submifllimeter Array (ALMA) provided the needed sensitivity and spatial resolution to resolve structures inside disks such as rings, gaps, and dust traps \citep{marel13,hltau}. These structures are ubiquitous in the star-forming regions studied in different ALMA programs (e.g., \citealt{Andrews2016, Long2018, Cieza2021}) and may reflect the presence of planets \citep{Zhang2015, Flock2015, Andrews2018}.

If those substructures are the consequence of the presence of planets, we should investigate the origin of planets in earlier evolutionary stages. The lack of sufficient mass to form the observed population of exoplanets in Class II disks \citep{Manara2018} and the fact that Class 0/I disks are likely more massive and solid-abundant \citep{Tychoniec2020-1, Tobin2020, Sheehan2017} support the scenario where planets are formed at earlier stages.

Dedicated high angular resolution observations millimeter/sub-millimeter wavelengths have revealed the presence of substructures in the dust disks surrounding some young Class 0/I sources \citep{Sheehan2017, Sheehan2018, Segura2020, Sheehan2020} but it remains a question whether these are ubiquitous and when they arise. To address this question, the ALMA Large Program 'Early Planet Formation in Embedded disks' \citep[eDisk]{eDisk} aims to study the origin of substructures systematically in Class 0/I disks with ALMA at 1.3 mm (Band 6) and a spatial resolution of 0\farcs04. The sample comprises twelve Class 0 and seven Class I protostars in nearby (d $<$ 200~pc) star-forming regions. The program aimed at detecting hints of disk substructures that may indicate the presence of planets or the presence of giant spirals as a byproduct of gravitational instabilities. 

The source GSS30 IRS3 was observed as part of the eDisk Large Program. It is part of the  L1688 molecular cloud, located at a distance of $\sim$138.4 $\pm$ 2.6 pc \citep{Ortiz-Leon18} based on the recent Gaia measurements. GSS30 IRS3 is a deeply embedded protostar \citep{Joergens09}, not detected at optical wavelengths, preventing a direct measurement of its distance from Gaia. Therefore, we assume that the distance to the source is similar to its host cloud. GSS30 IRS3 is classified as a Class 0 protostar, with a bolometric luminosity ($L\mathrm{_{bol}}$) of  1.7 L$_{\odot}$, and the bolometric temperature is 50 K \citep{eDisk}. GSS30 IRS 3 was first detected by \citet{Tamura91}  and later observed at 2.7 mm at higher-angular resolution in a focused paper on the three sources at GSS30 \citep{Zhang97}. The existence of a molecular outflow has been reported with single-dish observations taken at the James Clerk Maxwell Telescope \citep{White2015} and later with ALMA \citep{Friesen2018} in CO (2-1) at a resolution of $\sim$1\farcs5.

In this paper, we present observations of GSS30 IRS3 at a very high angular resolution (0\farcs056) in the continuum (at 1.3 mm) and ($\sim$0\farcs15) in line emission (CO isotopologues, H$_{2}$CO and c-C$_{3}$H$_{2}$) as part of the eDisk ALMA Large Program. The paper is structured as follows: we describe the observations in Section \ref{sec:p6_obs}. Then, we present the continuum, spectral line maps, and dust mass estimation in Section \ref{sec:p6_results}. Section \ref{sec:p6_analysis} presents the molecular outflow analysis and the estimation of the protostar's dynamical mass. We discuss the origin of the continuum asymmetries, twisted feature observed in the C$^{18}$O velocity map and the coexistence of a jet and a wide-angle outflow in Section \ref{Discussion}. Finally, we summarize the main results in Section \ref{sec:p6_summary}.

\begin{figure*}
\includegraphics[width=0.99\textwidth]{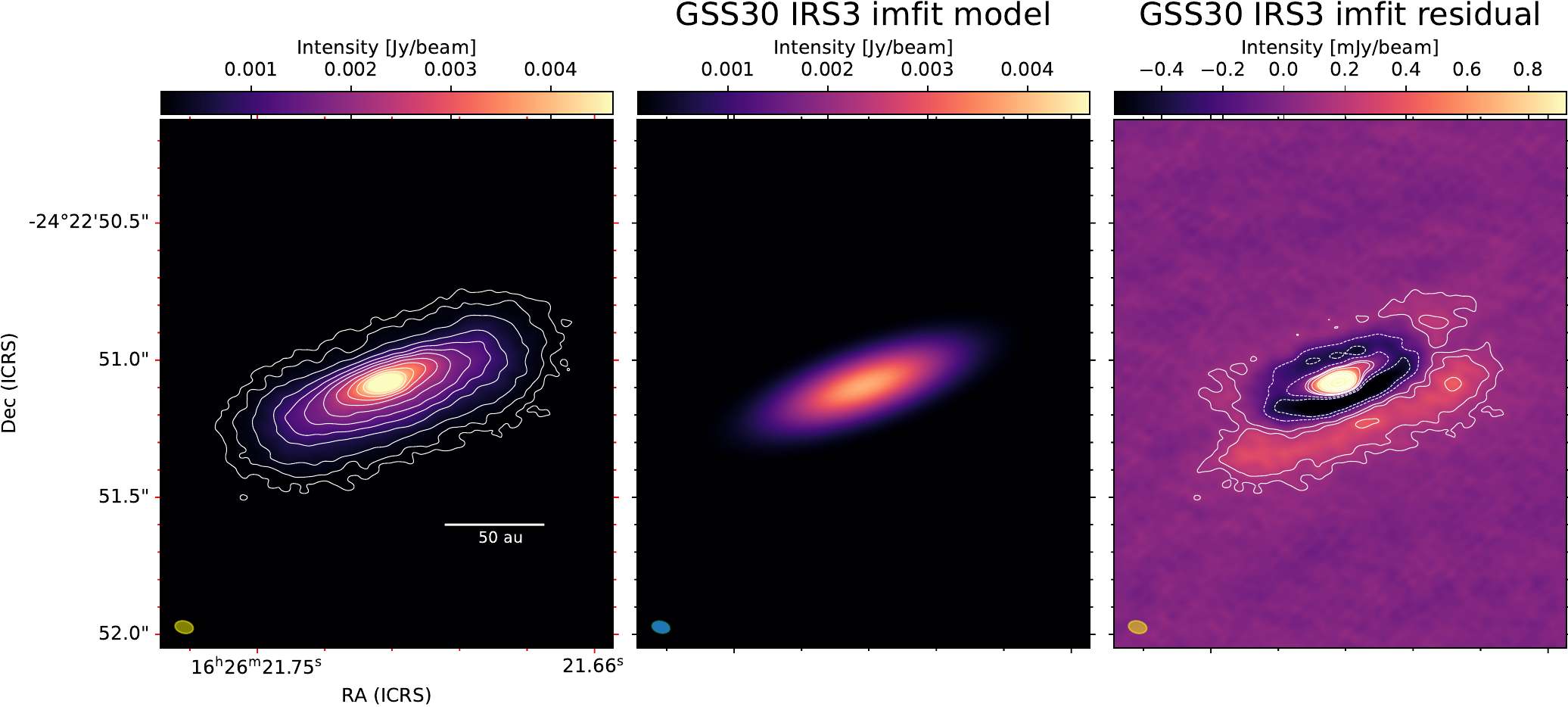}
\caption{Left: ALMA continuum image of GSS30 IRS3 at 1.3~mm. White contours are 5, 10, 20, 40, 60, 80, 100, 130, 160, 190, and 220 times the rms (1$\sigma$ = 18.5 $\mu$Jy beam$^{-1}$). Center: Gaussian fit model image. Right: Residual from the Gaussian fit model image. White contours are -20, -10, 5, 10, 20, 40, 60, 80, 100, 130, 160, 190, and 220 times the rms. Dashed lines represent negative contours. The beam size is represented by the filled ellipse in the bottom left corner of the three images.} \label{fig:continuum}
\end{figure*}

\begin{figure}
\includegraphics[width=0.45\textwidth]{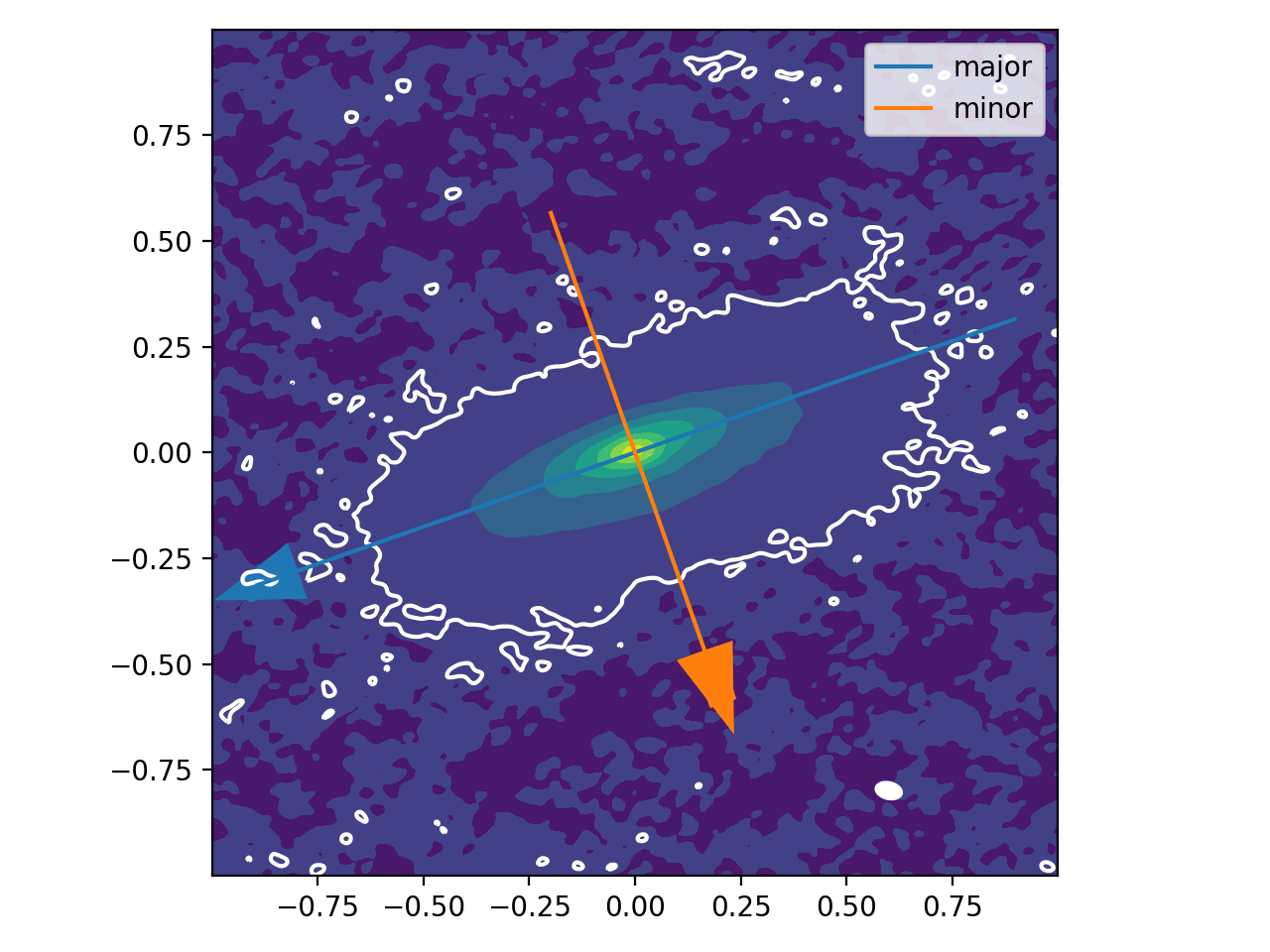}
\includegraphics[width=0.45\textwidth]{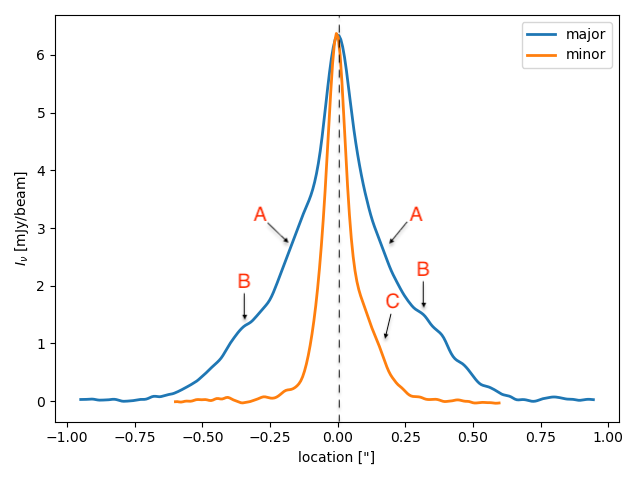}
\caption{Radial intensity profiles on the dust continuum emission. Top: 1.3 mm continuum image with major (blue) and minor (orange) axis overlaid showing the direction of the cuts. The white contours show the 3 $\sigma$ level. Coordinates of the central position are R.A.=16h26m21.715s, Dec= -24$\mathrm{^{o}}$ 22'51$\farcs$09, which is the peak position derived from the 2-D gaussian fitting. Bottom: Radial intensity profile cuts along the major and minor axes of the continuum disk; black arrows point to the bumps labeled with A, B for the major axis, and C for the minor axis. The dashed line shows the position of the continuum peak of the major axis derived from the 2-D gaussian fitting. Arrows in the top figure point to the positive part of the axis.} \label{fig:radial_profile}
\end{figure}

\begin{figure*} 
\includegraphics[width=0.95\textwidth]{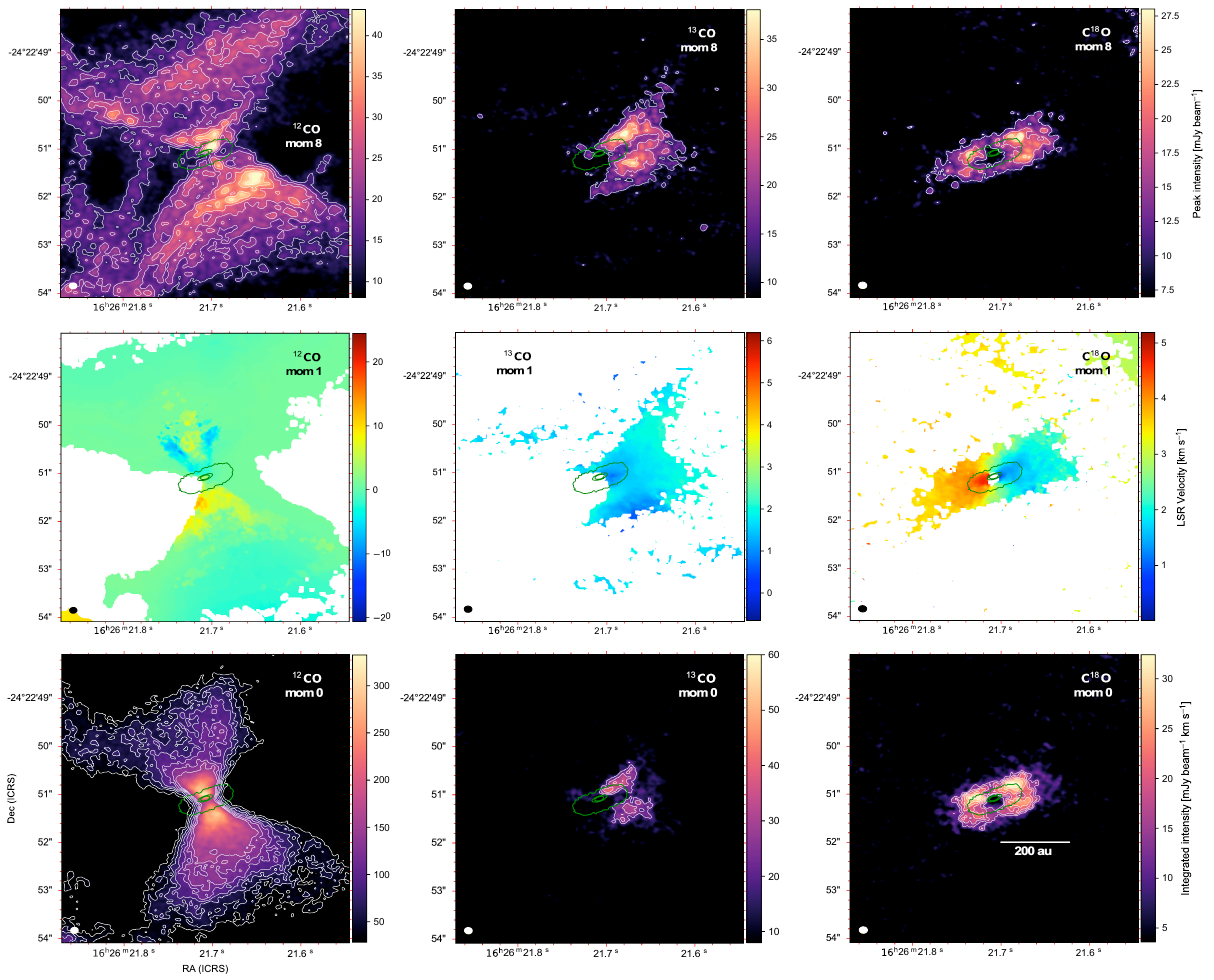}
\caption{Moment maps of the detected CO isotopologues. Left column shows $^{12}$CO (2-1) , middle column $^{13}$CO (2-1) , right column C$^{18}$O (2-1). The top row shows the peak intensity map, the center row shows the intensity-weighted velocity map, and the bottom row shows the integrated intensity map. The synthesized beam is represented as a filled ellipse on the bottom left side of every panel. Green contours represent the disk continuum emission with 10, 150 and 200 times the rms. White contours represent the 3, 5, 7, 9, 11, 13, 15 times the rms (1$\sigma$ = 9.5 mJy beam$^{-1}$ km s$^{-1}$) for the $^{12}$CO (2-1) moment 0 map; 5, 7, 9, 11, 13, 15 times the rms (1$\sigma$ = 2.7 mJy beam$^{-1}$) for $^{12}$CO (2-1) moment 8; 3, 5, 7, 9 and 11 times the rms for $^{13}$CO (2-1) moment 0 (1$\sigma$ = 4.0 mJy beam$^{-1}$ km s$^{-1}$) and 8 (1$\sigma$ = 4.2 mJy beam$^{-1}$); 3, 5, 7, 9, 11, 13, 15, 17 times the rms for C$^{18}$O (2-1) moment 0 (1$\sigma$ = 3.5 mJy beam$^{-1}$ km s$^{-1}$) and 8 (1$\sigma$ = 1.8 mJy beam$^{-1}$)}.\label{momentos}
\end{figure*}

\section{Observation and data reduction}
\label{sec:p6_obs}
GSS30 IRS3 was observed with the ALMA 12-m array in Band 6 as part of the eDisk ALMA Large Programme (project code: 2019.1.00261.L: PI N. Ohashi). A separate DDT program (2019.A.00034.S: PI J. Tobin) complements the eDisk data with shorter baselines for the same molecular lines for some sources. Observations toward GSS30 IRS3 from the Large Program were executed four times in October 2021 in configuration C43-8, while two executions were completed for the DDT program in June 2022 with configuration C43-5. Detailed information about the baseline length, number of antennas, precipitable water vapor, and calibrators are provided in \citet[Table 3]{eDisk}. 

The correlator was set up in dual-polarization mode using four basebands. The first baseband was divided into four spectral windows to detect C$^{18}$O (2-1), $^{13}$CO (2-1), H$_{2}$CO (3$_{2,1}$-2$_{2,0}$), and SO (6$_{5}$-5$_{4}$) with a spectral resolution of $\sim$0.17 km s$^{-1}$ and a bandwidth of 58.59 MHz. Basebands two and three were set up with a bandwidth of 1875 MHz to detect continuum. A spectral resolution of $\sim$1.3 km s$^{-1}$ was selected for the detection of several molecular lines (CH$_{3}$OH, SiO, DCN, H$_{2}$CO, and c-C$_{3}$H$_{2}$) and those lines were identified and masked when measuring the continuum. The fourth baseband has a bandwidth of 937.5 MHz (0.64 km s$^{-1}$ spectral resolution) to detect the continuum and the $^{12}$CO (2-1) line. The time on source for data from the Large Program was 2.75 h, and for the DDT program the time on source was 0.5 h. 

Calibration of the ALMA data was performed using the standard ALMA calibration pipeline. Then, we self-calibrated the data using the Common Astronomy Software Applications package \citep[CASA]{McMullin2007} version 6.2.1 and 6.4.1 for 2019.1.00261.L and 2019.A.00034.S programs, respectively. The pipeline version numbers should be supplied in addition to just the CASA number. Before performing the self-calibration, we first imaged the six executions and aligned the emission peaks using \textit{fixvis} and \textit{fixplanets}. After the alignment, we rescaled the flux in the UV plane to the execution on the 27 October. Then we proceeded with the self-calibration but using the rescaled non-aligned data, first with four phase-only cycles and two phase and amplitude cycles of the short-baseline data. Next, we performed joint self-calibration of the long and short baseline data and proceeded with four iterations of self-calibration in phase-only mode. Finally, the self-calibration solutions were applied to both the continuum and the spectral line data, and the task \textit{TCLEAN} was used to produce continuum and spectral line images after continuum subtraction. The final image is a combination of the short and long configurations. We adopted a Briggs robust parameter equal to 0 for the continuum image and 0.5 for the spectral line data as a compromise between sensitivity and resolution. We also adopted a Briggs robust parameter equal to 2 for the H$_{2}$CO and c-C$_{3}$H$_{2}$ spectral to enhance the sensitivity. The spectral line data were tapered at 2000 k$\lambda$ to bring out larger-scale structures. Primary beam correction was applied before inferring physical parameters from the images. The angular resolution of the continuum maps is 0$\farcs$056 (7.8 au at a distance of 138 PC), with a maximum recoverable angular scale of 2$\farcs$9, which is also the nominal value for Configuration 5 and a field of view of $\sim$22$''$. The absolute flux calibration uncertainty in Band 6 is estimated to be 10$\%$. A more detailed explanation of the reduction procedure can be found in \citet{eDisk}. 

\section{Results}
\label{sec:p6_results}

\begin{figure*}
\begin{center}
\includegraphics[width=0.95\textwidth]{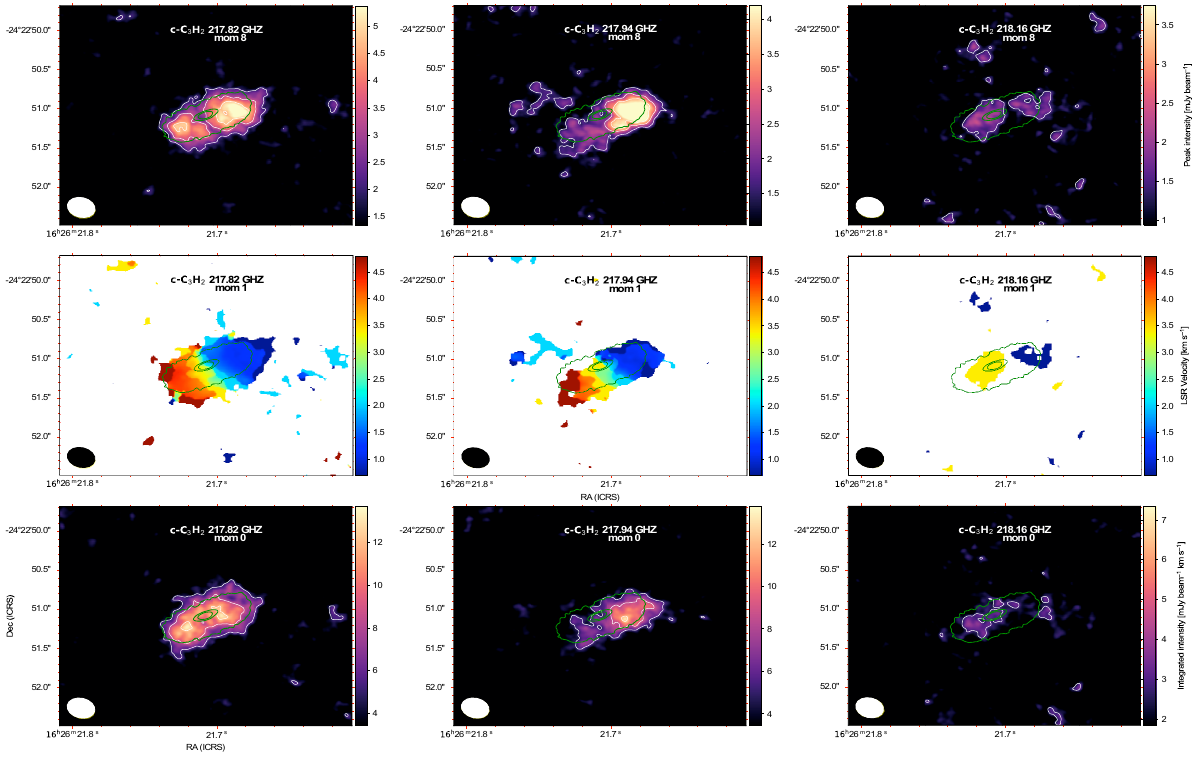}

\caption{Peak intensity (top row), intensity-weighted velocity map (center row), and integrated intensity (bottom row) maps of the detected c-C$_{3}$H$_{2}$ lines. The left column shows the images of the C$_{3}$H$_{2}$ transition at 217.822 GHz, the middle column for 217.94 GHz, and the right column for 218.16 GHz. The linear size is the same for all maps. The beam is represented as a filled ellipse on the bottom left. Green contours represent the disk continuum emission with 10, 150 and 200 times the rms given in Table \ref{Tab:gaussian_fit}. White contours represent the molecular emission at 3, 5, and 7 times the rms which is for the moment 0 maps 1$\sigma$ = 1.7 mJy beam$^{-1}$ km s$^{-1}$ (217.82 GHz), 1$\sigma$ = 1.7 mJy beam$^{-1}$ km s$^{-1}$ (217.94 GHz), and 1$\sigma$ = 0.9 mJy beam$^{-1}$ km s$^{-1}$(218.16 GHz) and for the moment 8 maps  1$\sigma$ = 0.6 mJy beam$^{-1}$ (217.82 GHz), 1$\sigma$ = 0.5 mJy beam$^{-1}$ (217.94 GHz), and 1$\sigma$ = 0.5 mJy beam$^{-1}$ (218.16 GHz)   }\label{momentos_c3h2}
\end{center}
\end{figure*}

\begin{figure*}
\includegraphics[width=0.95\textwidth]{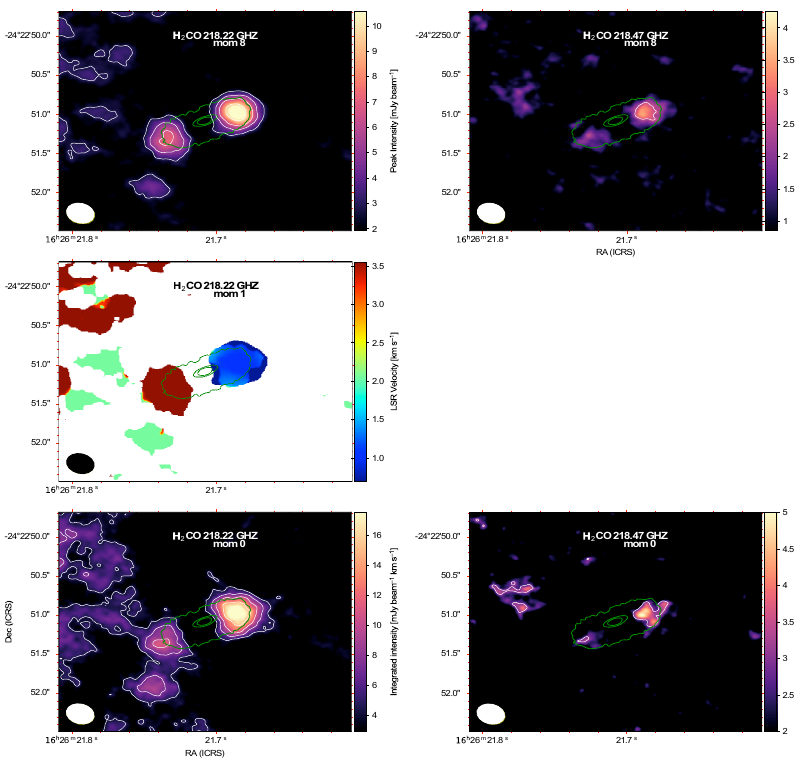}

\caption{Peak intensity (top row),  intensity-weighted velocity map (center row), and integrated intensity (bottom row) maps of the detected H$_{2}$CO lines. The left column shows the images of the H$_{2}$CO transition at 218.22 GHz, and the right column for 218.47 GHz. There is no intensity-weighted velocity map for 218.47 GHz transition because the emission is only located at 0.7 km s$^{-1}$. The linear size is the same for all maps. The beam is represented as a filled ellipse on the bottom left. Green contours represent the disk continuum emission with 10, 150 and 200 times the rms given in Table \ref{Tab:gaussian_fit}. White contours represent the molecular emission at 3, 5, and 7 times the rms which is for the moment 0 maps 1$\sigma$ = 1.5 mJy beam$^{-1}$ km s$^{-1}$ (218.22 GHz), and  1$\sigma$ = 1.0 mJy beam$^{-1}$ km s$^{-1}$ (218.47 GHz), and 1$\sigma$ = 1.0 mJy beam$^{-1}$ both moment 8 maps}.\label{momentos_h2co}

\end{figure*}

\subsection{Dust emission} \label{sec:dust_emission}
Figure \ref{fig:continuum} shows the dust emission at 1.3 mm of GSS30 IRS3, at an angular resolution of 0$\farcs$068$\times$0$\farcs$045 (Position angle of 76.61$\mathrm{^{o}}$), obtained using a robust parameter equal to 0. The continuum emission traces a disk-like structure around the central protostar with a high inclination angle. The deconvolved disk size is  0\farcs55$\times$0\farcs17 (76$\times$23 au) estimated from a two-dimensional Gaussian fitting (see Table \ref{Tab:gaussian_fit}). The best fit peak intensity is 3.87 $\pm$ 0.04 mJy beam$^{-1}$ at the Gaussian central position R.A.=16h26m21.715s, Dec= -24$\mathrm{^{o}}$ 22'51$\farcs$09. The flux density obtained from the Gaussian fitting is 123.6 $\pm$ 1.2 mJy. The continuum emission measured above a 5$\sigma$ contour has major and minor axes of 1\farcs43$\times$0\farcs62 ($\sim$198 $\times \sim$86 au), while the peak intensity and the flux density obtained by this method are 6.38 mJy beam$^{-1}$ and 133.7 mJy, respectively. Since the residuals from the Gaussian fitting show that the fit is far from perfect (see model and residual in Figure \ref{fig:continuum}), and also a fitting attempt using two (2-D) Gaussian components did not converge, we will adopt the values of the flux density and size inferred above the 5$\sigma$ contour in the analysis section. 
Assuming a geometrically flat and axisymmetric disk, we can estimate the disk inclination as $\cos i = b / a$, where $a$ and $b$  are the major and minor axes. The inclination derived using the deconvolved Gaussian fitting is 71.7 $\pm$ 0.3$\mathrm{^{o}}$ while using the 5$\sigma$ contour, we obtained 64.3 $\pm$ 1.5$\mathrm{^{o}}$, with the same position angle in both cases (109$\mathrm{^{o}}$). Given the poor quality of the Gaussian fitting, we will adopt a value of 64.3$\mathrm{^{o}}$ as the most reliable inclination estimation. We note that if the disk is geometrically thick this value should be treated as a lower limit.  Additionally, the dust disk exhibits a 'boxy' shape (see Section \ref{Discussion:asymmetries} for further discussions). 

Moreover, we also detected the dust emission of another source (GSS30 IRS1) in the field of view. See Section \ref{iras1} for further information.

\begin{table*}
\footnotesize
\caption{Physical parameters of the continuum map using \textit{imfit} task from CASA.}
\label{Tab:gaussian_fit}
\centering
\begin{tabular}{cccccc}
\hline \hline
rms & Flux density & Peak intensity & Major axis$^{\dagger}$ & Minor axis$^{\dagger}$ & Position angle$^{\dagger}$ \\

[$\mu$Jy beam$^{-1}$] & [mJy] & [mJy beam$^{-1}$] & [mas] & [mas] & [deg] \\
\hline
\hline  
18.5 & 123.6 $\pm$ 1.2 & 3.87 $\pm$ 0.04 & 549.5 $\pm$ 5.7 & 168.8 $\pm$ 1.8 & 109.36 $\pm$ 0.30 \\
\hline
\end{tabular}
\begin{flushleft}
Uncertainties are the ones provided by \textit{imfit}. $^{\dagger}$ FWHM deconvolved size
\end{flushleft}
\end{table*}

\subsection{Brightness asymmetries}
There are no apparent substructures in the disk surrounding GSS30 IRS3 (Figure \ref{fig:continuum}), such as clear spirals, gaps, or sharp rings, similar to those found in many more evolved Class II sources \citep{Andrews2018}. Instead, the image shows a striking asymmetry along the minor axis. To show that asymmetry more clearly, we performed a one-dimensional cut along the minor axis centered on the peak derived from the Gaussian fitting (Figure \ref{fig:radial_profile}). 
Additionally, along the major axis, there are two bumps on both sides of the disk, located at 0\farcs18  ($\sim$25 au) and 0\farcs36 ($\sim$50 au) East and West from the disk center, respectively. There is only one bump along the minor axis, which could be due to a flared structure of the disk.  The center of the bump is located at a deprojected distance of $\sim$20 au from the center of the disk. The radial brightness profile of the continuum emission is discussed in Sections \ref{Discussion:asymmetries} and \ref{bumps}.

\subsection{Dust disk mass}\label{sec:dust_disk_mass}
The dust disk mass surrounding GSS30 IRS3 is estimated under the assumption that the emission at 1.3 mm is optically thin and comes entirely from isothermal dust. We estimated the dust mass using the relationship from \citet{Hildebrand1983}:

\begin{equation}
      M =  \frac{S_{\nu} D^{2}}{ \kappa_{\nu}B_{\nu}(T_{d})},  
\end{equation}
where D is the distance to the source (138 pc),  $B_{\nu}(T_{d})$ is the Planck function at the observed frequency (225 GHz) at a dust temperature $T_{d}$, and $\kappa_{\nu}$(2.3 cm$^{2}$ g$^{-1}$) is the absorption coefficient adopted from \citet{Beckwith+1990}. $S_{\nu}$ is the flux density. We assumed  two different values of the dust temperature to estimate the dust disk mass. The first temperature value we used was a fixed value of 20 K, commonly assumed for Class II sources \citep{Pascucci16-1}. For the second temperature, we took into account the typical temperature of the disk based on the source's luminosity. We derived this temperature using the formula $\mathrm{T= 43(L_{bol}/L_{sun})^{0.25}}$ from \citet{Tobin2020}, which was obtained from a grid of radiative transfer models. For our specific case of GSS30 IRS3, this formula provides a temperature of 49 K.
Using these two temperature values, we calculated the dust disk mass using the flux density derived from the Gaussian fitting and the one calculated above the 5$\sigma$ contour level. For the Gaussian fitting, we obtained a dust disk mass of 69.5 $\pm$ 2.6 M$_{\Earth}$ and 23.87 $\pm$ 0.90 M$_{\Earth}$, respectively, for a temperature of 20 K and 49 K. The dust mass estimation when using the flux density above the 5$\sigma$ contour level provides values of  75.1 $\pm$ 2.9 M$_{\Earth}$ and 25.80 $\pm$ 0.99 M$_{\Earth}$ considering a temperature of 20 K and 49 K, respectively. The values that we derived are between 4.9 and 1.5 times more massive than the ones that were derived by \citet{Friesen2018}, however, their $\kappa_{\nu}$ value is much larger (7.4 cm$^{2}$ g$^{-1}$)  than the value used in this paper, so we associate the differences to the adopted absorption coefficient. Comparing to other Class 0 sources in the eDISK sample we find relatively similar values to R CrA IRS 5N (protostellar mass of 0.3 M$_{\odot}$) with a mass dust disk between 23.3 and 66.6 M$_{\Earth}$ \citep{Merel23} or L1527 (protostellar mass of 0.5 M$_{\odot}$) with a mass dust disk between 38.9 and 97.4 M$_{\Earth}$ \citep{Sharma2023}. Finally, we note that the derived dust mass might be a lower limit than the actual mass because the inclined disk could be optically thick.

\subsection{Gas emission}
We report the detection of several molecular species and transitions: $^{12}$CO (2-1), $^{13}$ CO (2-1), C$^{18}$O (2-1), c-C$_3$H$_2$ (blended lines 6$_{0,6}$-5$_{1,5}$ and 6$_{1,6}$-5$_{0,5}$), c-C$_3$H$_2$ (5$_{1,4}$-4$_{2,3}$),   c-C$_3$H$_2$ (5$_{2,4}$-4$_{1,3}$),  H$_{2}$CO (3$_{0,3}$-2$_{0,2}$) and H$_{2}$CO (3$_{2,2}$-2$_{2,1}$). The DCN (3-2), H$_{2}$CO  3$_{2,1}$-2$_{2,0}$, SO (6$_{5}$-5$_{4}$) and SiO (5-4), lines are not detected. The systemic velocity is 2.84 km s$^{-1}$ (which is derived from the C$^{18}$O (2-1) diagram fitting discussed in Section \ref{sec:SLAM_doble}). Moment maps of the detected molecular transitions are presented in Figures \ref{momentos}, \ref{momentos_c3h2} and \ref{momentos_h2co}, where we show integrated intensity (moment 0), mean velocity (moment 1), and peak intensity (moment 8) maps. A summary of the main properties of the detected lines including the angular resolution for each molecule is shown in Table \ref{tab:line_summary}, and an extensive description is provided in the following subsections.

\begin{figure*}
\centering 
\includegraphics[width=0.48\textwidth]{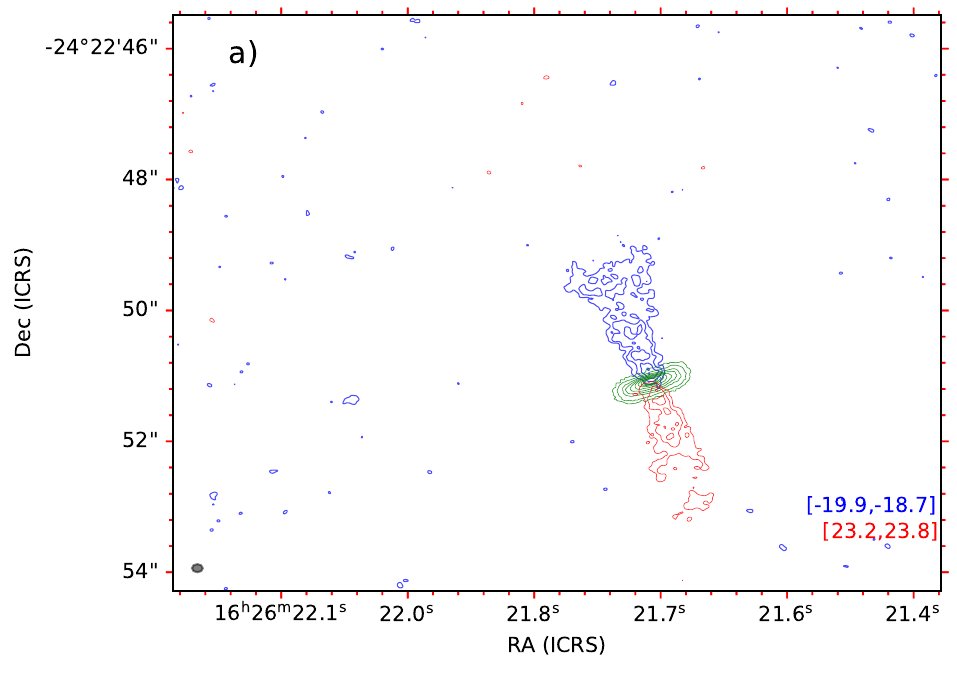}
\includegraphics[width=0.48\textwidth]{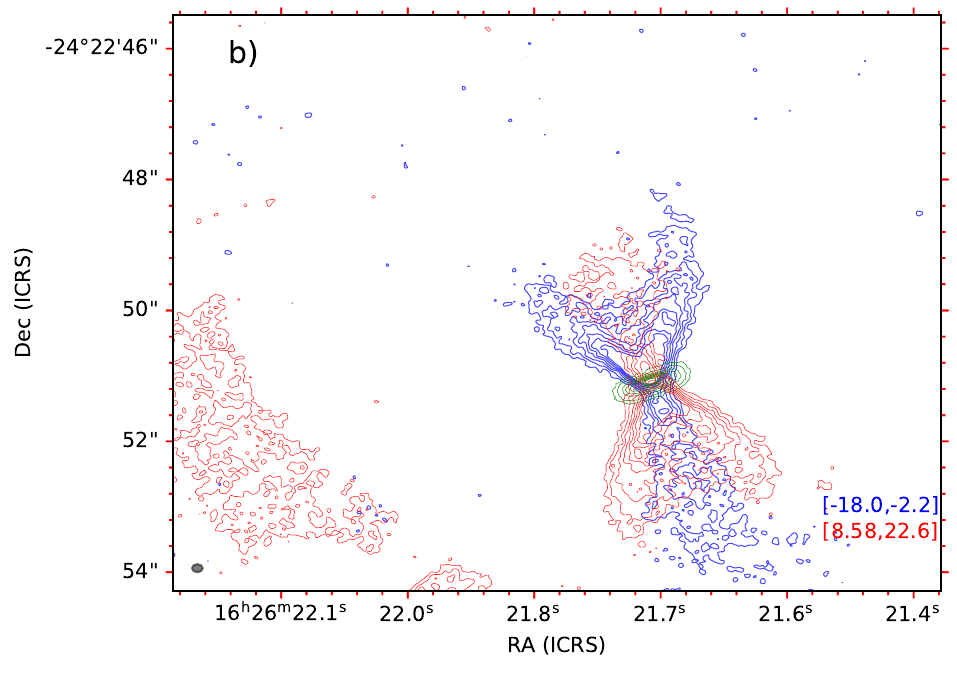} \\
\includegraphics[width=0.48\textwidth]{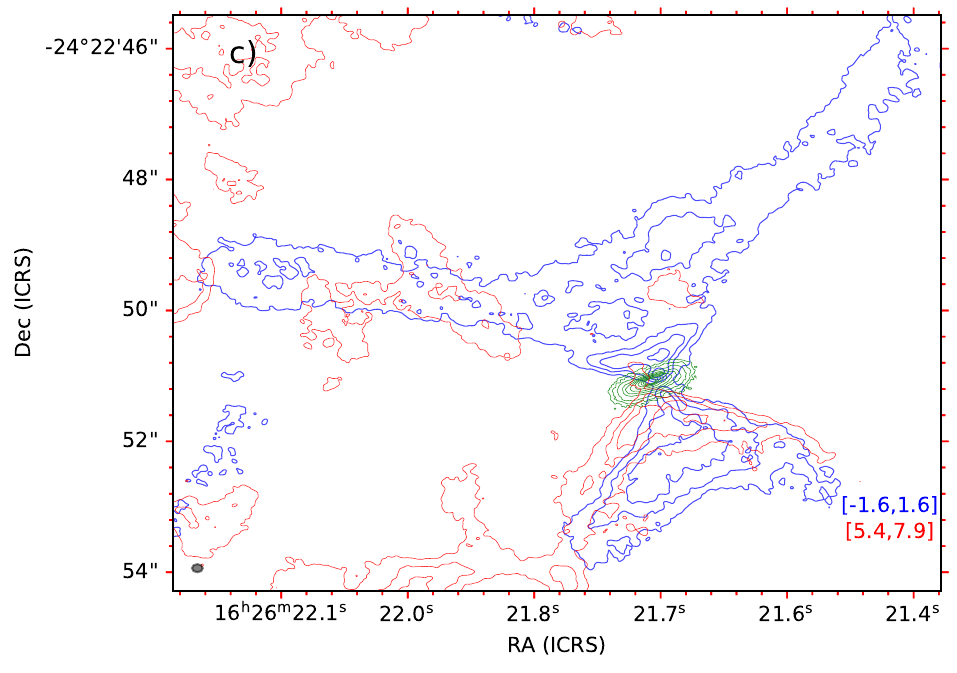} 
\includegraphics[width=0.48\textwidth]{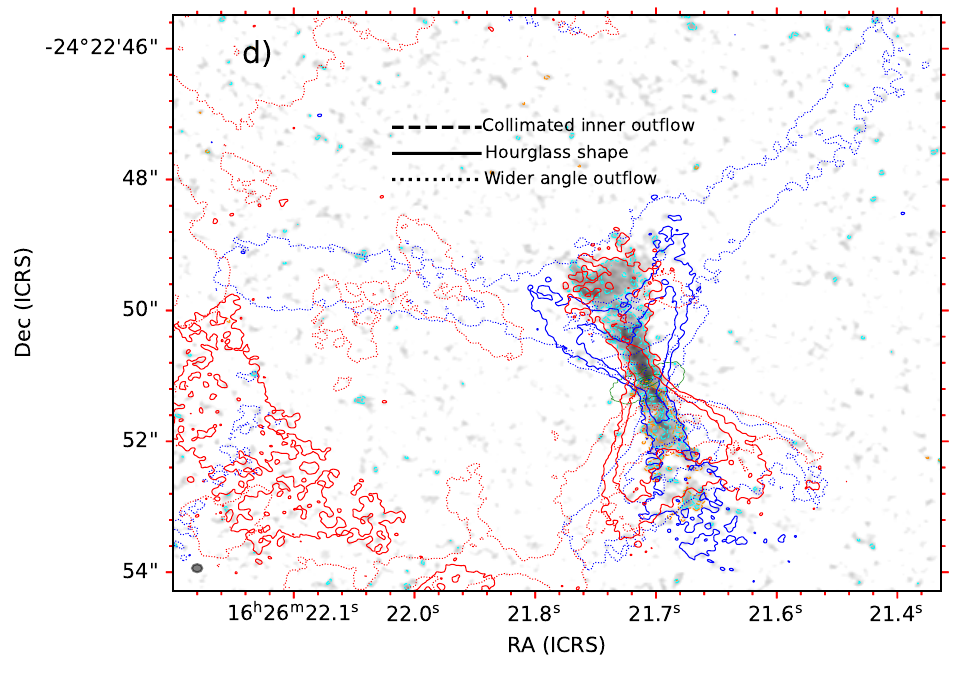}
\caption{$^{12}$CO (2-1) integrated emission maps (moment zero) at different velocity stages towards GSS30 IRS3. Green contours represent the dust disk in all the images. Blue contours show blueshifted emission and red contours show redshifted emission. a) b), and c) images show 3, 5, 7, 9, 11, 13, and 15 times the rms of the respective moment 0 map.  a) image of the inner high-velocity and narrow outflow component.  b) Close-up image of the hourglass shape intermediate velocity component of the molecular outflow. c) outer wide-angle low-velocity outflow emission. d) image is the combination of a) (dashed with orange and cyan colors), b) solid, and c) dotted using  3, and 7 times the rms. Grey image represents the inner high-velocity and narrow outflow. The local standard rest velocity of the cloud is $\sim$2.84 km s$^{-1}$.} \label{fig:wide_collimated}
\end{figure*}

\begin{table*}
\footnotesize
\caption{Main properties of the detected molecular lines}
\label{tab:line_summary}
\centering
\begin{tabular}{|cccccccc|}
\hline
 Frequency & Molecule & Transition & robust  & Beam size & $\Delta$v$^{a}$ &  Integrated intensity$^{b}$ & Peak intensity \\

 [GHz]       &    & &    & [$\arcsec$] & [km s$^{-1}$] & [Jy km s$^{-1}$] & [mJy beam$^{-1}$] \\

\hline 
217.822 & c-C$_3$H$_2$ & 6--5$^{c}$ & 2    & 0.359$\times$0.251 & 4.02 & 0.05630  $\pm$ 0.0061& 13.4 $\pm$ 2.2\\ 
217.940 & c-C$_3$H$_2$ & $5_{1,4}-4_{2,3}$ & 2 & 0.358$\times$0.251 & 4.02 & 0.0322 $\pm$ 0.0040 & 13.6 $\pm$ 2.2\\ 
218.160 & c-C$_3$H$_2$ & $5_{2,4}-4_{1,3}$ & 2  & 0.358$\times$0.251 & 2.78 & 0.0029  $\pm$ 0.0010& 4.0 $\pm$ 1.0\\ 
218.220 & H$_{2}$CO & $3_{0,3}-2_{0,2}$ & 2  &  0.358$\times$0.251 & 4.02 & 0.0591 $\pm$ 0.0060 & 19.4 $\pm$ 2.4\\
218.470 & H$_{2}$CO & $3_{2,2}-2_{2,1}$ & 2  & 0.358$\times$0.251 & 1.33 & 0.0270 $\pm$ 0.012& 5.06 $\pm$ 0.39\\ 
219.560 & C$^{18}$O & 2--1 & 0.5 & 0.171$\times$0.132 & 5.18 & 1.30 $\pm$ 0.13& 30.0 $\pm$ 3.9\\
220.398 & $^{13}$CO & 2--1 & 0.5 &  0.161$\times$0.126 &  6.85 &  1.11 $\pm$ 0.11& 42.0 $\pm$ 5.8\\ 
230.538 & $^{12}$CO & 2--1 & 0.5 & 0.158$\times$0.124 & 45.09 & 55.9 $\pm$ 5.6& 317 $\pm$ 33\\ 
\hline
\end{tabular} 
\begin{flushleft}
$^{a}$ Velocity width of the detected line measured with more than 3$\sigma$ detection.  \\
$^{b}$Integrated intensity over the whole emission area, obtained from a 3$\sigma$ contour over the moment 0 map. \\ 
$^{c}$ Two c-C$_3$H$_2$ lines are blended: ($6_{0,6}-5_{1,5}$ and  $6_{1,6}-5_{0,5}$). \\
Integrated intensity and peak intensity uncertainties consider a 10$\%$ absolute calibration uncertainty and the noise in the images.
\end{flushleft}
\end{table*}

\subsubsection{$^{12}$CO} \label{12co_description}
The $^{12}$CO (2-1) emission line traces primarily a bipolar molecular outflow centered at the position of GSS30 IRS3 and perpendicular to the dust disk, as was also previously reported by \citet{Friesen2018} using lower angular resolution and less sensitive observations. Our ALMA data reveal three different outflow components in morphology and velocity:  a collimated inner high-velocity component,  an intermediate velocity component with an hourglass or conical shape,  and a wider angle lower velocity component that is more prominent in the northern lobe which has a parabolic shape (see Figure \ref{fig:wide_collimated}). The velocity channel maps of the  $^{12}$CO (2-1) emission are presented in  Figure \ref{fig:channel_map_12co} and the intensity-weighted velocity map in Figure \ref{momentos}.  In Figure \ref{fig:JWST_image}, we plot the $^{12}$CO contours overlaid on a recent JWST image of the $\rho$ Ophiuchus region (P.I. Klaus Pontoppidan) where the H$_{2}$(0-0 S9) emission is tracing the outflow and delimiting very clearly the cavity wall beyond the molecular outflow. 

The collimated high-velocity component shows blueshifted emission to the northeast from the central source (spanning velocities between -19.9 and -18.7 km s$^{-1}$ in local standard rest velocity) and redshifted emission to the southwest (23.2 to 23.8 km s$^{-1}$; top left panel Figure \ref{fig:wide_collimated}). The JWST image reveals the presence of knots located at 4$\farcs$9 and 6$\farcs$7 from the center of the GSS30 IRS3 disk in the same direction of the blueshifted high-velocity component (right panel in Figure \ref{fig:JWST_image}). Correcting the outflow velocities from the inclination of the system, and considering that the inclination angle is a lower limit of the actual angle, the maximum deprojected velocity in the inner collimated high-velocity component is 52 km s$^{-1}$. It is important to note that the deprojected velocity derived is also a lower limit because  the inclination angle is a lower limit of the actual angle. This deprojected velocity falls within the expected velocity range observed in jets associated with Class 0 sources (100 $\pm$ 50 km s$^{-1}$, \citealt{Lee2015, Jhan2016, Podio2016, Yoshida2021}).

The intermediate velocity component of the outflow delineates an hourglass shape, and spanning velocities between  -18.0 and -2.2 km s$^{-1}$ for the blueshifted lobe and between 8.6 to 22.6 km s$^{-1}$ for the redshifted lobe. Both redshifted and blueshifted emission are present in both sides, as expected in outflows with a non-narrow opening angle and close to the plane of the sky  (see the top right panel in Figure \ref{fig:wide_collimated}). 
The wider angle low-velocity component of the molecular outflow shows blueshifted emission between -1.6 and 1.6 km s$^{-1}$m and redshifted emission between 5.4 and 7.9 km s$^{-1}$. It is formed by an extended but well-defined northeast arc mostly seen at blueshifted velocities and less extended southwestern lobe where both blueshifted and redshifted components overlay in the same area of the sky (see the bottom left panel in Figure \ref{fig:wide_collimated}).

We also report the detection of $^{12}$CO (2-1) in two other sources (GSS30 IRS1 and IRS2) that are in the ALMA field of view. GSS30 IRS1 observations will be presented in a future paper while the IRS2 detection is in Appendix \ref{iras2}.

\subsubsection{$^{13}$CO} \label{13co_description}
The $^{13}$CO (2-1) emission line, an optically thinner line due to its lower abundance than the $^{12}$CO (2-1) line, is detected at a peak S/N of 10. The velocity channel maps of the $^{13}$CO (2-1) emission line are presented in Figure \ref{fig:channel_map_13co}.
The blueshifted emission is found at velocities between -0.66 km s$^{-1}$ and  2.18 km s$^{-1}$, and it is stronger to the west of the central protostar. The redshifted emission is much fainter and found mainly to the east of the protostar. It spans velocities between 4.35  km s$^{-1}$ and 6.19  km s$^{-1}$. 

Two different faint structures are observed in the channel maps, tracing both the base of the northeast and southwest components of the molecular outflow and a gaseous disk-like structure surrounding the central protostar. The emission associated with the molecular outflow is only detectable between velocities of 0.35 km s$^{-1}$ and 2.35 km s$^{-1}$. The disk-like structure is detected between -0.66 and 2.18 km s$^{-1}$ (blueshifted component) and between 4.35 km s$^{-1}$ to 6.19 km s$^{-1}$ (redshifted component) velocities. The blueshifted component is more intense than the redshifted one. We plot the position-velocity diagram (PV) in Figure \ref{fig:c18o_fitting_single} to discriminate between redshifted emission associated with cloud or disk. Although it is a very faint emission, it is compatible with a rotating disk, which will be further discussed in Section \ref{sec:SLAM_doble}.

\subsubsection{C$^{18}$O} \label{sec:disco_c18} 
The C$^{18}$O (2-1) emission line is often considered optically thin and trace envelopes and disks, although it is only sometimes optically thick (e.g. \citealt{vanhoff18}). Blueshifted emission is detected at velocities ranging from 0.01 km s$^{-1}$ to 2.68 km s$^{-1}$ (see Figure \ref{fig:channel_map_c18o}). Redshifted emission spans velocities between 3.02 km s$^{-1}$ and 5.18 km s$^{-1}$. The size of the emission traced by the C$^{18}$O (2-1) emission above the 5$\sigma$ contour is 2\farcs3$\times$0\farcs93, with a position angle similar to the 109$\mathrm{^{o}}$ of the dust disk.  

\begin{figure}
\includegraphics[width=0.45\textwidth]{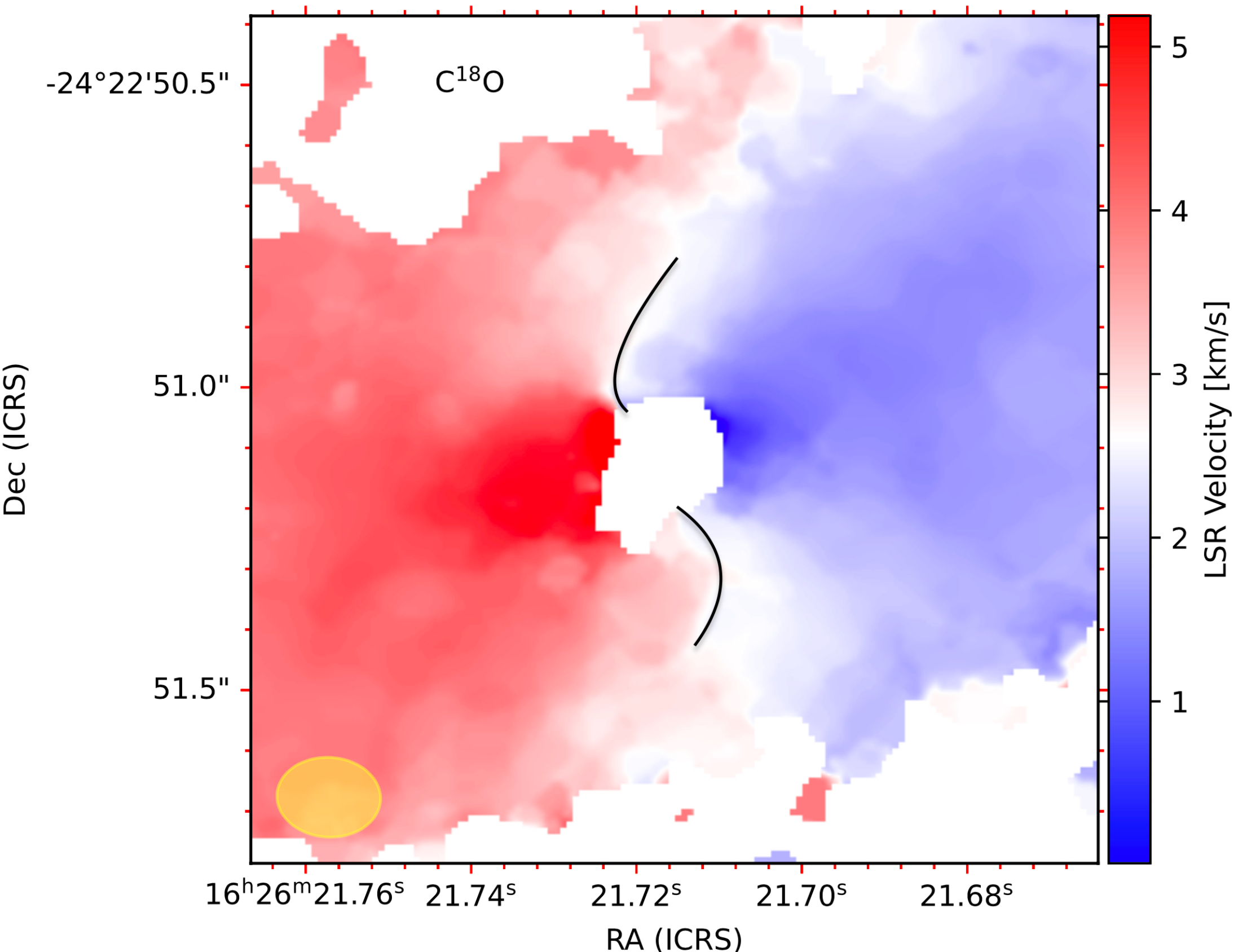}
\caption{Zoom in image of the velocity map of C$^{18}$O (2-1). The S shape is handwritten in black. The innermost white region does not show the systemic velocity.} \label{espiral}
\end{figure}

In GSS30 IRS3, C$^{18}$O (2-1) clearly traces a disk-like structure with an obvious redshifted and blueshifted pattern compatible with a Keplerian rotating disk (Figure \ref{momentos}). When studying the velocity map, we note a weak S-shape close to the systemic velocity (Figure \ref{espiral}). Its nature will be discussed in Section \ref{sec:warp_explanation}. 
The spectrum of C$^{18}$O (2-1) shows a double-peak profile typical of an inclined disk in Keplerian rotation.

\subsubsection{c-C$_{3}$H$_{2}$} 
 In Class 0 sources, c-C$_{3}$H$_{2}$ is usually located in the outflow cavity walls due to exposure to UV radiation from the central protostar \citep{Tychoniec2021}. On the other hand, in Class I sources, hydrocarbons such as c-C$_{3}$H$_{2}$ are located in the disk \citep{Tychoniec2021}. We detected the three targeted transitions of c-C$_{3}$H$_{2}$ at 217.822 GHz, 217.94 GHz, and 218.160 GHz. The first one has two lines blended ($6_{0,6}-5_{1,5}$ and  $6_{1,6}-5_{0,5}$) that are very close (<$<$m/s) for the kinematic analysis, and it is more intense than the other two detections (see Figure \ref{momentos_c3h2}). c-C$_{3}$H$_{2}$ at 217.822 GHz and 217.94 GHz shows emission between V$_{lsr}$ of 0.7 and 4.7 km s$^{-1}$ while the c-C$_{3}$H$_{2}$ at 218.16 GHz is only detected between 0.7 and 3.5 km s$^{-1}$. In GSS30 IRS3, the c-C$_{3}$H$_{2}$ emission is located in the same position as the dust disk Figure \ref{momentos_c3h2}). The association of the c-C$_{3}$H$_{2}$ detection is consistent with both the disk and the disk wind based on its positional information. Regarding kinematics, its velocities align with a rotation pattern. Therefore we tentatively attribute the association of c-C$_{3}$H$_{2}$ with the disk-like structure.

\subsubsection{H$_{2}$CO} 
Two transitions, H$_{2}$CO (3$_{0,3}$-2$_{0,2}$) and H$_{2}$CO (3$_{2,2}$-2$_{2,1}$), out of three, are detected. The transition H$_{2}$CO (3$_{0,3}$-2$_{0,2}$) at 218.22 GHz seems to be located in the same position as the external part of the dust disk and also exhibits a rotation pattern (Figure \ref{momentos_h2co}). H$_{2}$CO (3$_{0,3}$-2$_{0,2}$) emission is detected at V${lsr}$ values of 0.7 and 4.7 km s$^{-1}$. The transition H$_{2}$CO (3$_{2,2}$-2$_{2,1}$) is tentatively detected at the 3$\sigma$ level in one channel at a V$_{lsr}$ of 0.7 km s$^{-1}$.

\section{Analysis}
\label{sec:p6_analysis}

\begin{figure*}
\includegraphics[width=0.95\textwidth]{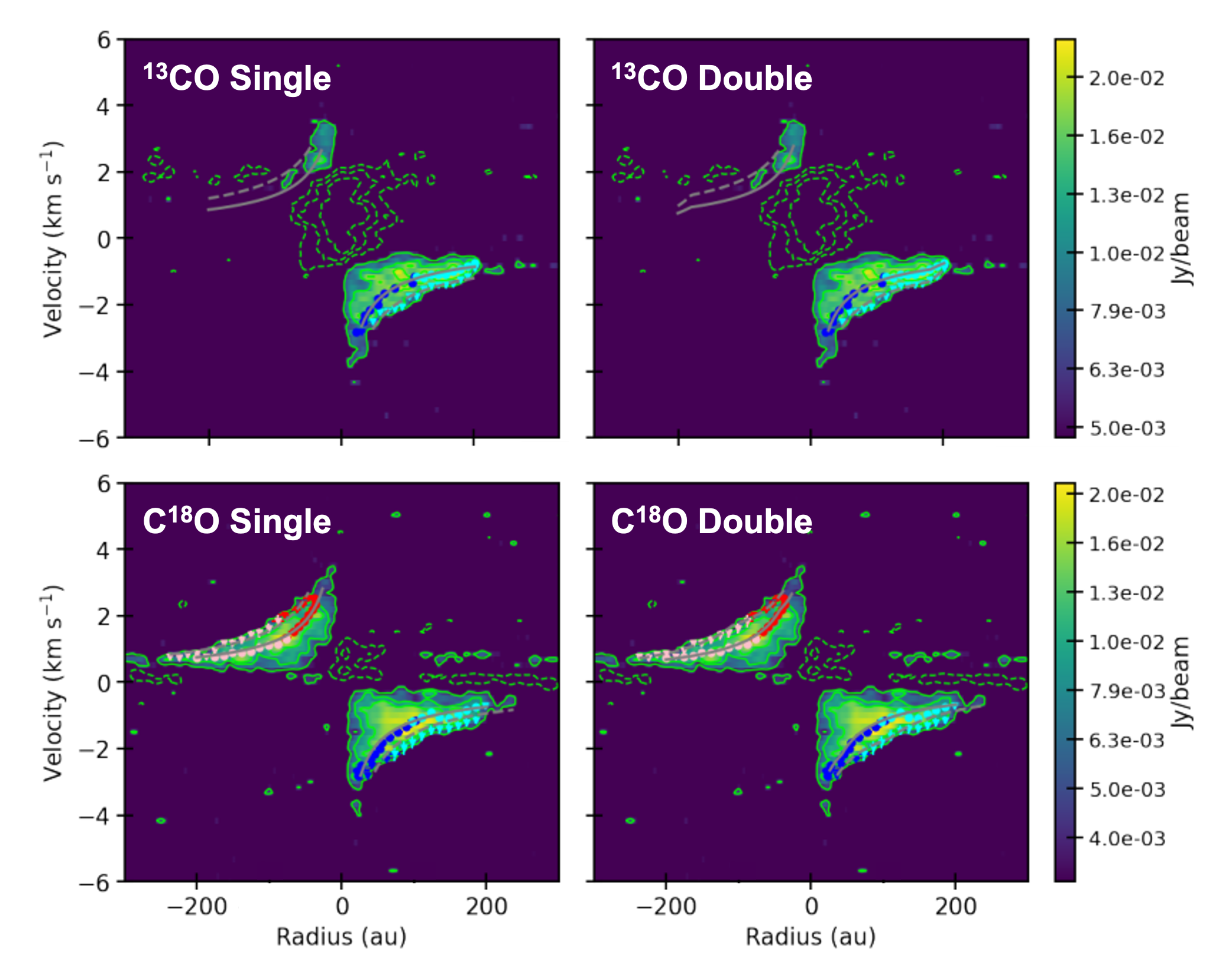} 
\caption{$^{13}$CO (2-1) and C$^{18}$O (2-1) position velocity diagrams. The top panel show $^{13}$CO (2-1) and the bottom panels show C$^{18}$O (2-1). All the diagrams were obtained along the disk major axis with a position angle of 109 degrees  (estimated from the 2-D gaussian fitting) and a width of 0\farcs17.  Circles represent the ridge component, while the triangles show the edge component using SLAM.  Solid and dashed curves represent fitting results obtained through the ridge and edge methods, respectively. Different colors show data points obtained along different directions: cyan and magenta represent data points obtained along the velocity axis, while blue and red show data points obtained along the position axis. The lime contours mark the emission at 3, 6 and 9 times the rms given in Table \ref{tab:line_summary}.Further details of the fitting procedure can be found in \citet{yusuke_aso_2023}. The left column is for the single power-law fit, while the right column is for the double power law fit.}\label{fig:c18o_fitting_single}\label{fig:c18o_fitting_double}
\end{figure*}

\subsection{Disk rotation}\label{sec:SLAM_doble}
The C$^{18}$O (2-1) emission line traces a disk-like structure exhibiting indications of rotation (Section \ref{sec:disco_c18}). It is important to investigate the nature of the rotation, in particularly whether it is Keplerian. If it is Keplerian rotation, the dynamical mass of the central star can be derived. In order to investigate the rotation curve of the C$^{18}$O (2-1) emission, we will use a PV diagram. The C$^{18}$O (2-1) line is the most suitable tracer for this analysis because the disk-like structure is clearly detected with good S/N. This is consistent with being optically thinner than the $^{13}$CO (2-1) line, which partially traces the gaseous disk in GSS30 IRS3. In any case, we analyze the rotation curve of both molecular transitions to compare the obtained results. For computing the position-velocity diagram, we use a PV diagram cut of 0\farcs17 width, that corresponds to the size of the major axis of the synthesized beam, along the major axis of the C$^{18}$O (2-1) and  $^{13}$CO (2-1) disk. We adopt a disk position angle value of 109$\mathrm{^{o}}$ estimated in Section \ref{sec:dust_emission}. 

The SLAM code\footnote{\url{https://github.com/jinshisai/SLAM}} \citep[Spectral Line Analysis/Modeling]{yusuke_aso_2023} is used to analyze the rotation curve of GSS30 IRS3. The process is based on two main steps; the first step is to obtain a 1-D intensity profile to derive the edge and ridge data points. Ridge points are defined as the intensity-weighted mean position or the center in the 1-D intensity profile from the Gaussian fitting. The edge points or radius is the outermost position at a certain S/N level. The second step is fitting a power-law (single or double) function, that is defined as a positive value, to the derived representative data as a function of velocity or radius using the Markov-Chain Monte Carlo method. Whether it is a function of radius or velocity depends on along which axis the 1D profile for ridge/edge calculations is obtained see \citet{yusuke_aso_2023} criteria to use the v$_{i}$-r$_{i}$ data points as a function of radius or velocity. Note that the edge and ridge data points are independently fitted. Thus, the double power law implies two components having different rotation laws, such as a Keplerian disk and an infalling envelope conserving its specific angular momentum, where the break radius ($r\mathrm{_{b}}$) is the boundary radius between a Keplerian disk and a rotation/infalling envelope, and where the break velocity ($v\mathrm{_{b}}$) is the velocity at ($r\mathrm{_{b}}$). We also derived the reduced chi-squared ($\chi^{2}$) for each fit in order to compare all the results. A more detailed explanation of the SLAM methodology for the eDisk papers can be found in \citet{eDisk}.

The SLAM code can fit the systemic velocity as a free parameter. We ran a first test to derive the systemic velocity as 2.84 km s$^{-1}$ for the C$^{18}$O (2-1) emission, then we checked this value with the C$^{18}$O (2-1) spectra and fixed this value for the following test until we derived the final dynamical mass. 

We tested a single-power law (bottom-left panel in Figure \ref{fig:c18o_fitting_single}) and double-power law fit (bottom-right panel in Figure \ref{fig:c18o_fitting_double}) to the C$^{18}$O (2-1) PV diagram. The single power-law fittings give power-law indices (p$\mathrm{_{in}}$) of $\sim$0.71 $\pm$ 0.03 for the edge method and $\sim$0.65$\pm$0.02 for the ridge method as shown in Table \ref{tab:13co_slam_results}. These values are between 0.5 (Keplerian rotation) and 1 (constant specific angular momentum), suggesting that the observed rotation may not be explained by only Keplerian rotation and may also include rotation in an infalling envelope. To investigate such possibility, double power-law fittings were also performed; the fitting results with the edge method give a power-law index (p$\mathrm{_{in}}$) of $\sim$0.45$\pm$0.08 and dp is the difference of the index between the inner and outer parts that is $\sim$0.65$\pm$0.15 with r$\mathrm{_{b}}$ of 107$^{+26}_{-19}$ au as shown in Table \ref{tab:13co_slam_results_doble}, meaning that the power-law index for the outside of r$\mathrm{_{b}}$ is $\sim$1.1. These results suggest that the observed rotation may be explained as a combination of Keplerian rotation and rotation conserving its specific angular momentum. This kind of transition of the rotation between a Keplerian disk and an infalling envelope was observed before in some other protostars as well \citep[e.g.,][]{Ohashi14-1, Aso15-1, Aso17-1, Sai20}. The ridge method was not converging using p$\mathrm{_{in}}$ as a free parameter, so we fixed it at 0.5 under the assumption that there is a Keplerian disk and we derived a mass of  0.26  $\pm$ 0.07 M$_{\odot}$. Then, the mass derived from the single power-law fit is 0.27 $\pm$ 0.07 M$_{\odot}$ and  0.35 $\pm$ 0.09 M$_{\odot}$ for the double power law fit as the average of the ridge and edge methods because the edge and the ridge methods over- and under-estimate the stellar mass by a factor that depends on the resolution \citep{Maret2020}.

For the $^{13}$CO (2-1) line, we only considered the blueshifted component of the emission for the fit, since the redshifted component is very faint. The single-power law fit (top-left panel in Figure \ref{fig:c18o_fitting_single}) shows a lower value of p$\mathrm{_{in}}$ than the C$^{18}$O (2-1) single-power fit. The double-power law fitting is not converging. The single-power fit is returning a mass at the break radius (M$_\mathrm{b}$) of 0.24 $\pm$ 0.01 M$_{\odot}$. The edge method results were discarded because we are not obtaining stable solutions after several iterations. Additional details about the fitting are described in Appendix \ref{app:c18o_fitting}.

The best $\chi^{2}$, ignoring the value of ridge double-power law because p$\mathrm{_{in}}$ is fixed, is the one obtained from  C$^{18}$O (2-1) single power fit ridge method, however the p$\mathrm{_{in}}$ is not close to the expected value of a Keplerian disk, so we should discard this value. Then, the second best  $\chi^{2}$ value is the one obtained from C$^{18}$O (2-1) edge double-power law fit. Therefore, we adopt a mass of GSS30 IRS3 is 0.35$\pm$0.09 M$_{\odot}$.

\subsection{Characterization of the molecular outflow}\label{char_outflow}
In this subsection, we analyze deeper the three different components of the molecular outflow described in Section \ref{12co_description}. 

The most collimated inner high-velocity outflow component is extended to a maximum deprojected length of 2\farcs3 (1.5$\times$10$^{-3}$pc). The maximum deprojected extension of the intermediate velocity component with an hourglass shape is 3\farcs3, while the maximum deprojected length of the wider angle lower velocity component is 8\farcs3 and 2\farcs7  for the northeast and southwest arc-like structures, respectively. The size of the different components was measured using a bisecting line from the center of the disk continuum emission to the maximum extension of each component, considering the 5$\sigma$ contour.

The most collimated inner high-velocity outflow component exhibits an opening angle of $\sim$34 degrees. On the other hand, the intermediate velocity component with an hourglass shape shows an opening angle of $\sim$70 degrees. For the wider angle low-velocity outflow component, we estimate values of $\sim$120 degrees and $\sim$100 degrees for the northeastern and southwestern lobes, respectively. These values were measured by hand using a protractor over the flux-integrated map (Figure \ref{fig:wide_collimated}). Table \ref{tab:geometrical_parameters} summarizes all these values.  

Given that our observations were not designed nor optimized for studying extended molecular outflows and that we are likely filtering structures larger than $\sim$2\farcs0, we derived the dynamical parameters of the outflow as a first characterization in Appendix \ref{dynamical_params}, however, the results in Table \ref{tab:dynamical_parameters} must be considered cautiously and valid only for the most compact emission of the outflow.

\begin{table}
\footnotesize

\caption{Results from the SLAM fitting to the position velocity diagram using a single power law.} \label{tab:13co_slam_results}
\centering
\begin{tabular}{ccc}
 & $^{13}$CO (2-1) & C$^{18}$O (2-1) \\
\hline \hline
Edge method \\ 
p$\mathrm{_{in}}$ & 0.601 $\pm$ 0.010 & 0.714  $\pm$ 0.029\\
M$_{b}$ [M$_{\odot}$]  & 0.430 $\pm$ 0.004 & 0.316  $\pm$ 0.007\\
R$_{b}$ [au] & 120.5 $\pm$ 1.2 & 111.0 $\pm$ 2.5\\
$\chi^{2}$ & - & 8.2 \\
\hline
Ridge method\\
p$\mathrm{_{in}}$ & 0.57 $\pm$ 0.010 & 0.655  $\pm$ 0.023 \\
M$_{b}$ [M$_{\odot}$]  & 0.244 $\pm$ 0.002 & 0.211  $\pm$ 0.004\\
R$_{b}$ [au] & 70.45  $^{+0.72}_{-0.75}$ & 73.5 $\pm$ 1.5 \\
$\chi^{2}$ & 25 & 2.4  \\ 
\hline
\end{tabular}
\begin{flushleft}
p$\mathrm{_{in}}$ is the power-law index,  $\chi^{2}$ is the reduced chi-squared of the fit. M$_{b}$ is the protostellar mass derived given the rotation velocity. R$_{b}$ is the radius where M$_{b}$ was measured assuming Keplerian rotation.
\end{flushleft}
\end{table}

\begin{table}
\footnotesize

\caption{Results from the SLAM fitting to the position velocity diagram using a double power-law} \label{tab:13co_slam_results_doble}
\centering
\begin{tabular}{ccc}
& $^{13}$CO (2-1)  &  C$^{18}$O (2-1) \\
\hline \hline
Edge method \\ 
R$_{b}$ [au]   & 181 $\pm$ 48 &108 $^{+25}_{-20}$ \\ 
v$_{b}$ [km/s] & 1.30 $\pm$ 0.39 &1.71 $\pm$ 0.24 \\
p$\mathrm{_{in}}$ & 0.54 $\pm$ 0.13 &0.460  $\pm$ 0.076\\
dp  & 2.2 $\pm$ 1.8 &0.65 $^{+0.19}_{-0.13}$ \\
M$_{b}$ [M$_{\odot}$] & 0.42 $\pm$ 0.28 & 0.44  $\pm$ 0.16 \\
$\chi^{2}$ & - &5.6\\

\hline
Ridge method\\
R$_{b}$ [au]   & 181 $\pm$ 5.5& 47 $^{+8}_{-7}$  \\ 
v$_{b}$ [km/s] &  0.934 $\pm$ 0.020 &2.03  $\pm$ 0.18 \\
p$\mathrm{_{in}}$ & 0.565 $\pm$ 0.012 & 0.5  (fixed) \\
dp & 1.52 $\pm$ 0.56  &0.22  $\pm$ 0.04  \\
M$_{b}$ [M$_{\odot}$] & 0.220 $\pm$ 0.012  & 0.26  $\pm$ 0.06  \\
$\chi^{2}$ & 23 & 2.0 \\
\hline
\end{tabular}
\begin{flushleft}
p$\mathrm{_{in}}$ is the power-law index, R$_{b}$ is the radius where the power-law index changes and v$_{b}$  is the velocity at  R$_{b}$. M$_{b}$ is the protostellar mass derived given the rotation velocity at the break radius assuming Keplerian rotation and $\chi^{2}$ is the reduced chi-squared of the fit.\end{flushleft}
\end{table}

\begin{table*}
\footnotesize
\caption{Geometric parameters of the three  molecular outflow components  (northeast and southwest component) deprojected for an inclination of 64.3$^{o}$.} \label{tab:geometrical_parameters}
\centering
\begin{tabular}{ccccccc}
& \multicolumn{2}{c}{Collimated} & \multicolumn{2}{c}{Hourglass shape} & \multicolumn{2}{c}{Wider angle} \\
& \multicolumn{2}{c}{high-velocity} & \multicolumn{2}{c}{} & \multicolumn{2}{c}{low-velocity} \\
\cmidrule(l){2-3} 
\cmidrule(l){4-5}
\cmidrule(l){6-7}
&  Blue & Red & Blue & Red & Blue & Red \\

\hline
Size [$\arcsec$] &2.3 & 2.4 & 3.3 & 2.8 & 8.3 & 2.7 \\
V$\mathrm{_{max}}$ [km s$^{-1}$] & 52.3 & 48.4 & 48.1 & 45.5 & 10.2 & 11.9 \\ 
$\tau_{dyn}$ [yr] & 30 & 32 & 45 & 40 & 535 & 147 \\
Opening angle [$^{o}$] & 38 & 34 & 70 & 67 & 117 & 97 \\
\end{tabular}
\end{table*}

\section{Discussion} \label{Discussion}
\subsection{Asymmetric Intensity Distribution along the minor axis of the Dust Disk}
\label{Discussion:asymmetries}
The eDisk results has shown the presence of asymmetric intensity distribution along the disk minor axis in several sources \citep[e.g.,][]{eDisk, Takakuwa23, Lin23, Sai23, Han2023, Miyu2023, Merel23, Gavino2023}. The asymmetries found in the minor axis in these eDisk subsample are generally explained due to a combination of optically-thick emission and disk flaring, as proved in the radiative transfer studies of \citet{Lin23} and \citet{Takakuwa23}. As a result, dust is not settled onto the midplane but is distributed vertically, which, in some cases, can create a dark lane. This dark lane observed in edge-on disks has been described as a ``hamburger" shape \citep{Lee2017, Galvan-Madrid18-1} .

In the case of the orientation of the GSS30 IRS3 disk, the northeastern region corresponds to the far side of the disk. This conclusion is based on the peak intensity position, which exhibits a slight skew from the geometric center and towards the northeast, as determined by fitting an ellipse to a 5$\sigma$ contour level. Moreover, this is supported by the orientation of the collimated high-velocity molecular outflow, where the blueshifted emission points towards the northeast. This alignment provides a rationale for the observed brightness asymmetry in the minor axis between the two sides of the disk. Additionally,  the dust disk 'boxy' shape (Figure \ref{fig:continuum}) can be potentially explained by the truncated edge of a flared disk. 

Consequently, the asymmetry in the minor axis can be attributed to a combination of disk flaring and optically thick emission.

\subsection{The origin of the bumps along the major axis}\label{bumps}
A subset of eDisk sources reported the presence of brightness asymmetry or bumps along the major axis of the protostellar disks \citep{Sai23, Merel23, Miyu23}. In the case of GSS30 IRS3, two bumps are detected along the disk major axis. 
To confirm the authenticity of the bumps and eliminate the possibility of them being spurious artifacts from the cleaning process, we fit the radial intensity profile in the uv plane using \textit{Frank} (see Appendix \ref{Analysis:asymmetries}). \textit{Frank} \citep{Jennings2020} first deproject the visibilities to correct the disk inclination and uses the Fourier-Bessel series to reconstruct the radial brightness profile. The bumps are located at a radius of $\sim$0\farcs19 ($\sim$26 au) and the other located at $\sim$0\farcs36 ($\sim$50 au) from the center of the disk. The positions of the identified bumps, closely coincide with those obtained from the radial intensity profile in the image plane. This alignment strongly rules out the possibility of being artifacts. However, we emphasize that our interpretation is only limited to discarding spurious signals in the cleaning process given the \textit{Frank} limitations when dealing with a flared disk. 

We discuss two main possible explanations for the origin of the bumps. First, there could be a real substructure within the dusty disk that will likely evolve into a more well-defined structure over time. Alternatively, the bumps might be the result of the temperature distribution due to optically thick continuum emission. To test the second possibility, we compare the dust brightness temperatures at the bumps' locations with the expected midplane temperature at the same location in a passively heated and flared disk.

\begin{figure}
\includegraphics[width=0.45\textwidth]{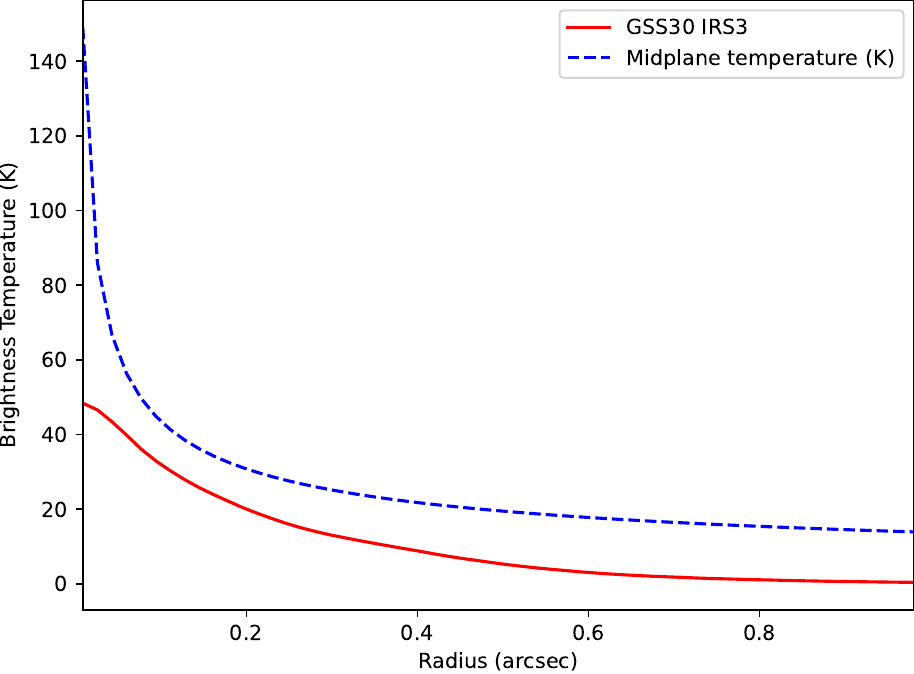}
\caption{Azimuthal brightness distribution versus the radius for GSS30 IRS3 (red continuous line) and the theoretical midplane temperature (dashed blue line).} \label{tmidr}
\end{figure}

To obtain the midplane temperature, we employ the equation T$_{\mathrm{mid}}$(r) = ($\phi$ L$_{*}$/8$\pi$r$^{2}\sigma_{\mathrm{SB}}$)$^{0.25}$  \citep{Chiang97, Alessio98, Dullemond01}, where r is the radius (0\farcs19 and 0\farcs36), $\sigma_{\mathrm{SB}}$ denotes the Stefan-Boltzmann constant, L$_{*}$ is the bolometric luminosity of the central protostar (1.7 L$_{\odot}$) and $\phi$ corresponds to the assumed flaring angle of 0.02 radian \citep{Huang2018}. The predicted midplane temperature is 27.8 K at a radius of 0\farcs19 and 20.2 K at a radius of 0\farcs36. However, the observed brightness temperature at  0\farcs19  and 0\farcs36 are 18.9 K and 9.6 K, respectively. The observed brightness distribution along the major axis of the disk is compared with the estimated midplane temperature distribution in Figure \ref{tmidr}, which shows that the observed brightness temperature is lower than the estimated midplane temperature. This suggests that the 1.3 mm continuum emission is not completely optically thick, as we would expect the observed brightness temperature to be similar to the actual physical temperature. It is, however, advisable to exercise caution in interpreting the observed brightness temperature distribution as an accurate representation of the actual radial brightness temperature distribution of the disk. This is due to the fact that the disk is inclined at a significant angle, which introduces a number of complications. In an edge-on disk, the brightness temperature is measured at a point where the optical depth along the line of sight becomes unity. Consequently, the distance of the measurement point from the center is not necessarily equivalent to the corresponding project distance from the center.

Future continuum observations at lower frequencies with ALMA that trace optically thinner dust emission could help to validate our findings. Also, a more detailed radiative transfer modeling, which is beyond the scope of this paper, might further elucidate the true nature of these bumps and their relationship with the substructures observed in more evolved sources.

\begin{table*}
\footnotesize
\caption{Inclination corrected dynamical properties of the blueshited and redshifted emission of the three molecular outflow components} \label{tab:dynamical_parameters}
\centering
\begin{tabular}{|lcccccc|}
\hline 
& \multicolumn{2}{c}{Collimated} & \multicolumn{2}{c}{Hourglass shape} & \multicolumn{2}{c}{Wider angle} \\
& \multicolumn{2}{c}{high-velocity} & \multicolumn{2}{c}{} & \multicolumn{2}{c}{low-velocity}    \\
\cmidrule(l){2-3} 
\cmidrule(l){4-5}
\cmidrule(l){6-7}
Properties & Blueshifted & Redshifted & Blueshifted & Redshifted & Blueshifted & Redshifted  \\
\hline

Outflow mass [M$_{\odot}$] & 4.2$\times10^{-7}$ & 2.0$\times10^{-7}$ & 9.4$\times10^{-6}$ & 1.1 $\times10^{-5}$ & 3.4$\times10^{-5}$ & 6.1$\times10^{-6}$ \\
Mass-loss rate [M$_{\odot}$ / yr] & 1.4$\times10^{-8}$ & 6.3$\times10^{-9}$ & 2.1$\times10^{-7}$ & 2.7$\times10^{-7}$ & 6.4$\times10^{-8}$ & 4.1$\times10^{-8}$ \\
Momentum [M$_{\odot}$ km s$^{-1}$] & 2.2$\times10^{-5}$ & 9.8$\times10^{-6}$ & 4.5$\times10^{-3}$ & 4.8$\times10^{-4}$ &  3.5$\times10^{-4}$ & 7.2$\times10^{-5}$ \\
Energy [erg] & 1.1$\times10^{40}$ &  4.7$\times10^{39}$ & 2.2$\times10^{41}$ & 2.2$\times10^{41}$ & 3.5$\times10^{40}$ & 8.4$\times10^{39}$ \\
Luminosity [L$_{\odot}$] & 3.0$\times10^{-3}$ & 1.1$\times10^{-3}$ & 3.7$\times10^{-2}$ & 4.3$\times10^{-2}$ & 5.2$\times10^{-4}$ & 4.5$\times10^{-4}$ \\
Mechanical force [M$_{\odot}$ km s$^{-1}$ yr$^{-1}$] &  7.4$\times10^{-7}$ & 3.0$\times10^{-7}$ & 1.0 $\times10^{-5}$ & 1.2$\times10^{-5}$ &  6.5$\times10^{-7}$ & 4.8$\times10^{-7}$ \\
\hline
\end{tabular}
\end{table*}

\subsection{Kinematics and the origin of the S-shape in the C$^{18}$O line emission}
\label{sec:warp_explanation}
In this subsection, we discuss the origin of the twisted feature observed in the C$^{18}$O (2-1) velocity map (Section \ref{sec:disco_c18}). Similar S-shape or twisted velocity map close to the systemic velocity are usually associated with warped disks in young stellar objects \citep{Rosenfeld12, Facchini18, Zhu19}. There are several possible explanations for the origin of these features: the existence of a misaligned binary \citep{Terquem1993,Papaloizou1995, Fragner2010, Nixon2013, Facchini18}, a fast radial infall attributed to the presence of a giant planet, brown dwarf, or low-mass companion \citep{Rosenfeld2014}, the effect of a stellar fly-by \citep{Dai15, Kurtovic2018, Cuello19}, or the transition from a Keplerian disk to an infalling envelope \citep{Aso15, Sai20}. The first two theories are usually applied to more evolved young stellar objects. 

In the eDisk sample there is another source with a twisted kinematic map\citep[Oph IRS63]{Flores2023}. The C$^{18}$O (2-1) and SO (6-5) moment 1 maps showed an S-shaped feature which is explained by infalling motion. In the case of GSS30 IRS3 observations we are also detecting an infalling envelope according to the double-power law fit from SLAM (see Section \ref{sec:SLAM_doble})
Although the most likely explanation is an infalling envelope onto the disk, the twisted kinematic feature alone does not provide sufficient evidence for further analysis. On the other hand,  we are unable to confirm that the origin of the S-shape is due to the presence of an undetected close binary (within $\sim$8 au) or an embedded companion. Also, we do not observe indications of a recent star flyby. Future modeling, incorporating hydrodynamic simulations may be able to explain the origin of the S-shape; however, such modeling is beyond the scope of this initial analysis.

\subsection{Coexistence of a jet and a wide-angle outflow}
Accretion and ejection are fundamental processes in star formation, shaping the evolution of young stars and their environment. The relation between ejection and accretion processes is not completely understood, but the accretion of gas and dust that fall from the protoplanetary disk´s inner edge onto the young stellar object drives the ejection of material. Ejection of material removes the excess of angular momentum through two different types of structures: a collimated component at high velocity and a wider-angle component at lower velocities.  The collimated component or jet propagates into the surrounding molecular cloud, forming bow shocks that push the gas and produce outflow shells in the jet´s vicinity \citep{Raga1993}. Different knots are usually detected along the jet direction as a signal of episodic material ejection. On the other side, the wide-angle component is a radial disk wind that blows the ambient cloud, forming expanding shells \citep{Li-1996b, Lee2000}. Therefore, the jet and wind angle disk-driven structures are able to explain the mass loss process in star formation with different theories. Interestingly, \citet{Raben2022} presented simulations demonstrating that jet and wide-angle disk-wind driven ejections could coexist, expecting the simultaneous detection of collimated and wide-angle outflow components during the early stages of star formation

The $^{12}$CO observations of GSS30 IRS3 reveal three distinct components: a collimated outflow which axis is aligned with several knots seen in molecular hydrogen  (Figure \ref{fig:JWST_image}), a conical hourglass shape and a wide-angle, low-velocity component with a parabolic morphology (Figure \ref{fig:wide_collimated}). The analysis of the conical outflow, described in Section \ref{12co_description}, suggests that it is produced by a jet. While the blueshifted lobe of the wider angle outflow clearly exhibits emission shifting away from the central object as velocity increases, this is consistent with the predictions of the wind-driven shell model \citep{Lee2000}. 

The coexistence of collimated jets/outflows and slow wider-angle molecular outflows has been previously detected in young protostars at early stages, specifically in two Class I sources in HH 46/47 \citep{Zhang2019} and DG Tauri B \citep{Valon2020, Valon2022} or more recently in several sources \citep{Federman23}. Compared with those sources, GSS30 IRS3 exhibits very distinguishable components, even with observations that were not initially designed to study molecular outflows. Moreover, this is one of the few sources that the coexistence of a jet and a wide-angle outflow has been clearly detected in a young Class 0 protostar in the submm regime, making this source one of the best objects for future detailed studies of the accretion-ejection processes at the early stages of star formation.

\section{Conclusions}
\label{sec:p6_summary}
We present high-resolution ALMA observations at 1.3 mm of the Class 0 protostar GSS30 IRS3, conducted as part of the eDisk ALMA program at a spatial resolution of $\sim$8 au. Our observations targeted the continuum emission as well as several molecular species, including $^{12}$CO (2-1), $^{13}$CO (2-1), C$^{18}$O (2-1), and c-C$_{3}$H$_{2}$ (blended lines 6$_{0,6}$-5$_{1,5}$ and 6$_{1,6}$-5$_{0,5}$ 217.822 GHz), c-C$_3$H$_2$ (5$_{1,4}$-4$_{2,3}$ 217.940 GHz), c-C$_3$H$_2$ (5$_{2,4}$-4$_{1,3}$ 218.16 GHz), H$_{2}$CO (3$_{0,3}$-2$_{0,2}$ 218.22 GHz), and H$_{2}$CO (3$_{2,2}$-2$_{2,1}$ 218.47 GHz). We summarize our findings as follows: \\

1. We detect dust continuum emission tracing a disk-like structure. The dust disk size is 1\farcs43$\times$0\farcs62 ($\sim$198 $\times \sim$86 au) with an inclination of 64.3 $\pm$ 1.5$\mathrm{^{o}}$ and a  position angle of 109$\mathrm{^{o}}$. The mass of the dust disk in GSS30 IRS3 is estimated to range between 23.87 $\pm$ 0.90 M$_{\Earth}$ and 75.1 $\pm$ 2.9 M$_{\Earth}$ depending on the adopted T$_{dust}$. \\
 
2. The dust disk of GSS30 IRS3 does not exhibit clear substructures such as spirals. We report the presence of an asymmetry in the minor axis, which can be attributed to an optically thick emission and disk flaring. Additionally, we note the detection of two bumps located at 26 au and 50 au from the disk center along the major axis. The origin of these bumps remains unclear, and we propose two main hypotheses: they could indicate embedded real substructures within the disk, or that we are tracing the temperature distribution. \\ 

3. The C$^{18}$O (2-1) emission traces a Keplerian disk and the infalling rotating envelope, allowing us to derive a dynamic mass of 0.35$\pm$0.09 M$_{\odot}$ of the central protostar with a disk gas size radius between 47 and 107 au. 

4. Our $^{12}$CO (2-1) high-resolution and sensitive observations reveal the coexistence of a jet and a wide-angle outflow. \\

\begin{acknowledgements}
    We thank the referee, Tien-Hao Hsieh, for the helpful comments and suggestions on this manuscript. We would like to thank all the ALMA staff supporting this work. This paper makes use of the following ALMA data: ADS/JAO.ALMA$\#$2019.1.00261.L, 2019.A.00034.S ALMA is a partnership of ESO (representing its member states), NSF (USA) and NINS (Japan), together with NRC (Canada), MOST and ASIAA (Taiwan), and KASI (Republic of Korea), in cooperation with the Republic of Chile. The Joint ALMA Observatory is operated by ESO, AUI/NRAO and NAOJ. J.J.T. acknowledges support from NASA XRP 80NSSC22K1159.  The National Radio Astronomy Observatory is a facility of the National Science Foundation operated under cooperative agreement by Associated Universities, Inc.
    N.O. acknowledges support from National Science and Technology Council (NSTC) in Taiwan through the grants NSTC 109-2112-M-001-051, 110-2112-M-001-031, 110-2124-M-001-007, and 111-2124-M-001-005.
    J.K.J. acknowledges support from the Independent Research Fund Denmark (grant No. 0135-00123B). C.W.L. is supported by the Basic Science Research Program through the National Research Foundation of Korea (NRF) funded by the Ministry of Education, Science and Technology (NRF-2019R1A2C1010851), and by the Korea Astronomy and Space Science Institute grant funded by the Korea government (MSIT; Project No. 2022-1-840-05).
    Z.-Y.L. is supported in part by NASA NSSC20K0533 and NSF AST-2307199 and AST-1910106.
    L.W.L. acknowledges support from NSF AST-2108794.
    J.P.W. acknowledges support from NSF AST-2107841.
    W.K. was supported by the National Research Foundation of Korea (NRF) grant funded by the Korea government (MSIT) (NRF-2021R1F1A1061794).
ZYDL acknowledges support from NASA 80NSSCK1095, the Jefferson Scholars Foundation, the NRAO ALMA Student Observing Support (SOS) SOSPA8-003, the Achievements Rewards for College Scientists (ARCS) Foundation Washington Chapter, the Virginia Space Grant Consortium (VSGC), and UVA research computing (RIVANNA). S.T.  is supported by JSPS KAKENHI grant Nos. 21H00048  and 21H04495, and by NAOJ ALMA Scientific Research grant No. 2022-20A.
    H.-W.Y.\ acknowledges support from the National Science and Technology Council (NSTC) in Taiwan through the grant NSTC 110-2628-M-001-003-MY3 and from the Academia Sinica Career Development Award (AS-CDA-111- M03).
    IdG acknowledges support from grant PID2020-114461GB-I00, funded by MCIN/AEI/10.13039/501100011033.
    ASM thanks Laura P\'erez and Anibal Sierra for their valuable help and insights on using the Frank. 
\end{acknowledgements}
\bibliographystyle{aa.bst} 
\bibliography{bibliographyv3}{}

\begin{appendix}
\section{Channel maps} 
This section shows the channel maps for $^{12}$CO (Section \ref{12co_description}), $^{13}$CO (Section \ref{13co_description}) and C$^{18}$O (Section \ref{sec:disco_c18})
\begin{figure*}[h!]
\includegraphics[width=0.85\textwidth]{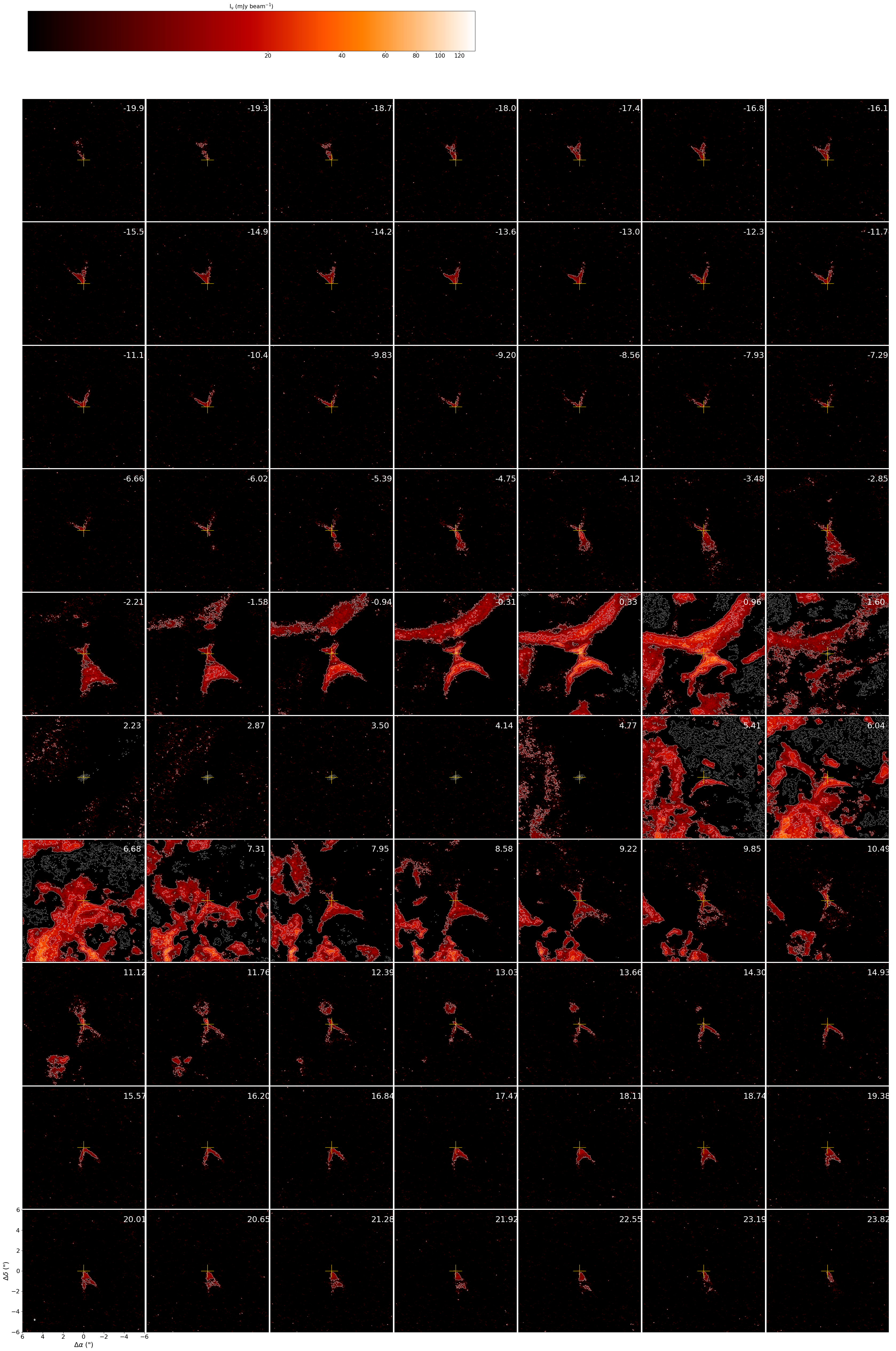}
\caption{Channel maps of the $^{12}$CO (2-1) emission in GSS30 IRS3. Emission is detected above 5$\sigma$ from  -19.9 km s$^{-1}$ to 23.8 km s$^{-1}$. The color scale is stretched using the inverse hyperbolic sine function. The lowest contour emission corresponds to 5$\sigma$ (1$\sigma$ corresponds to 1.00 mJy beam$^{-1}$), with increments of 10$\sigma$. The field of view of the map is $12\arcsec\times12\arcsec$.} \label{fig:channel_map_12co}
\end{figure*}

\begin{figure*}[h!]
\includegraphics[width=0.90\textwidth]{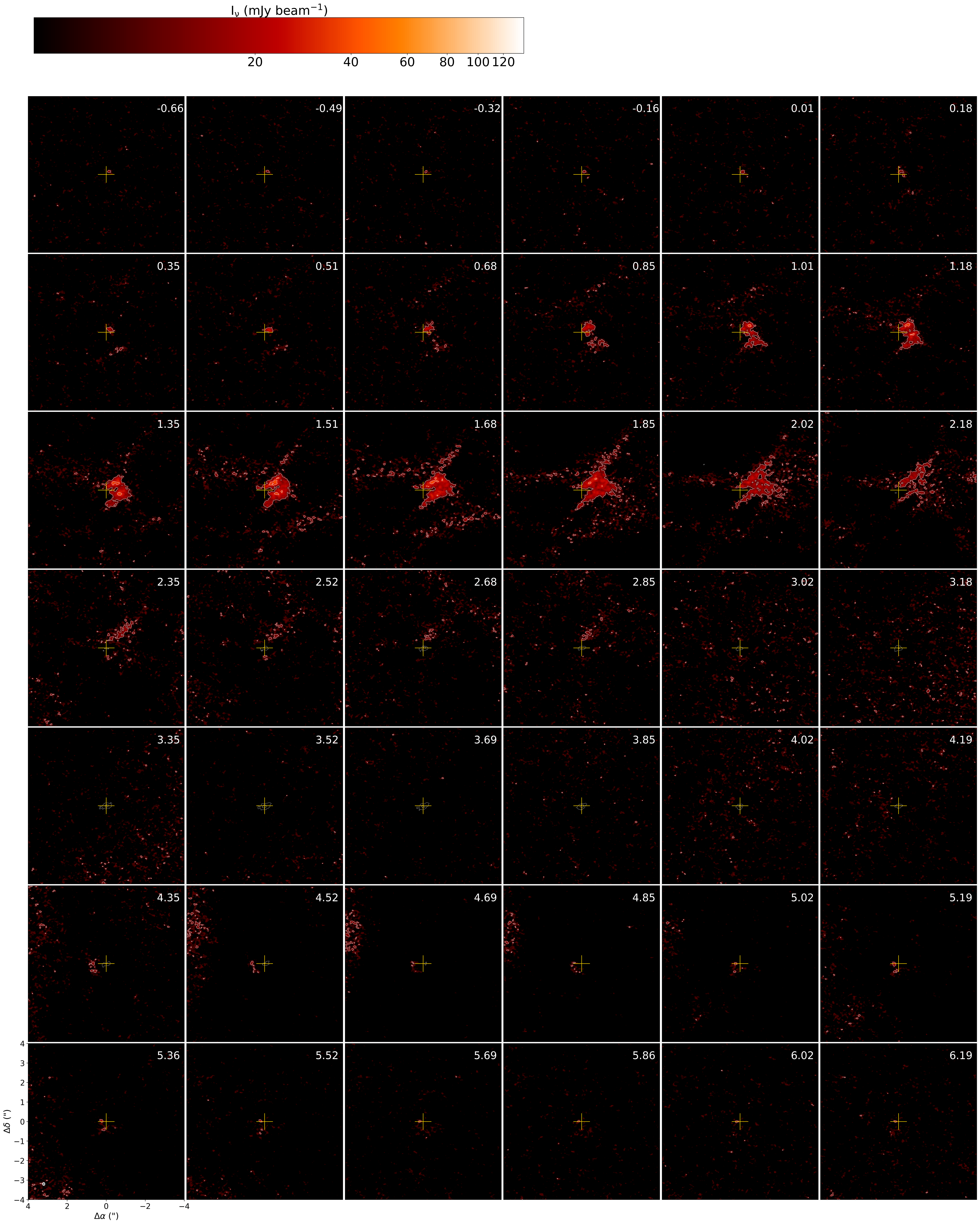}
\caption{Channel maps of the $^{13}$CO (2-1) emission in GSS30 IRS3. Emission is detected above 5$\sigma$ from -0.66 km s$^{-1}$ to 6.19 km s$^{-1}$. The color scale is stretched using the inverse hyperbolic sine function. The lowest contour emission corresponds to 5$\sigma$ (1$\sigma$ corresponds to 2.10 mJy beam$^{-1}$), with increments of 10$\sigma$. The field of view of the map is $8\arcsec\times8\arcsec$.} \label{fig:channel_map_13co}
\end{figure*}

\begin{figure*}[h!]
\includegraphics[width=0.90\textwidth]{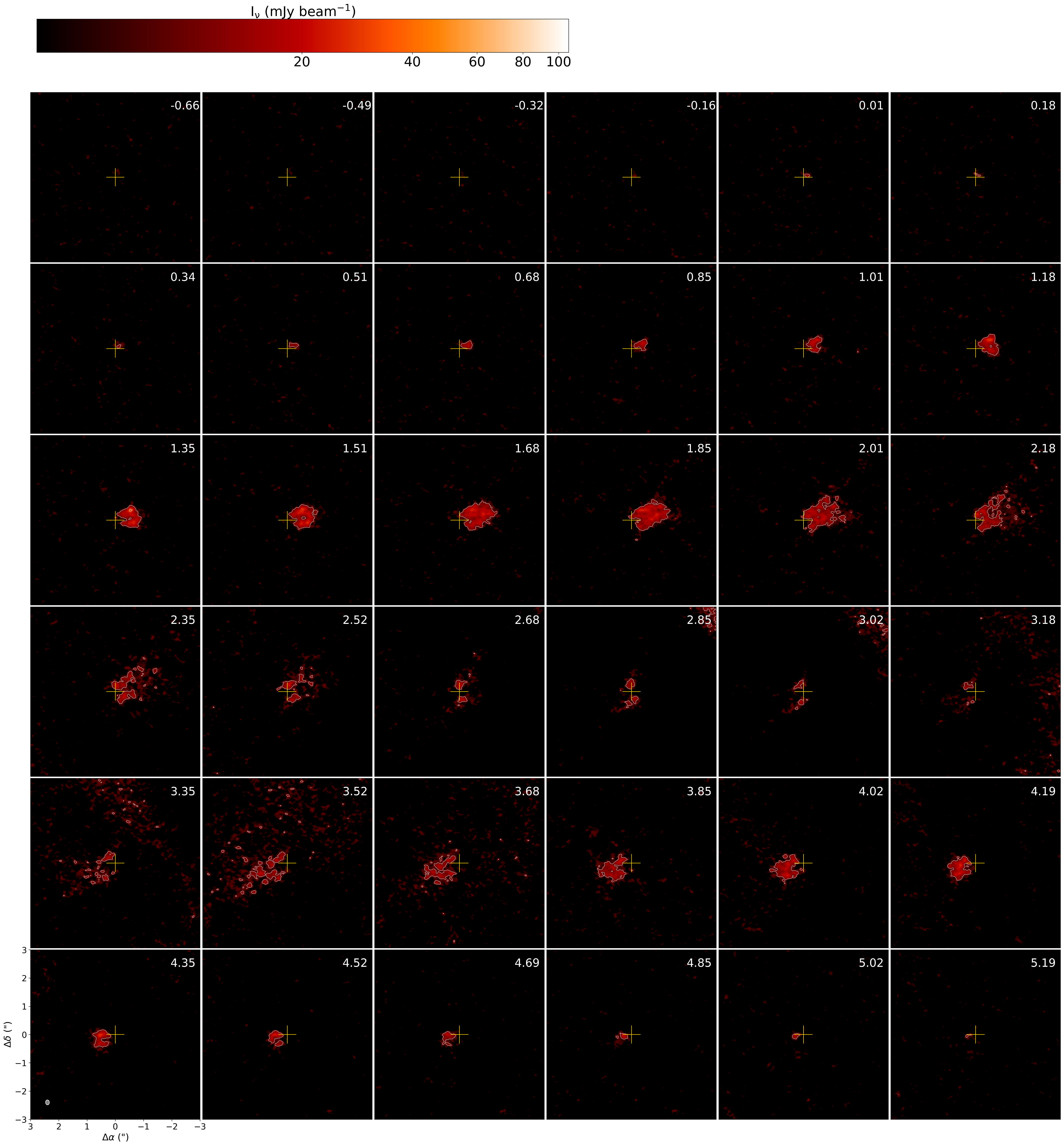}
\caption{Channel maps of the C$^{18}$O (2-1) emission in GSS30 IRS3. Emission is detected above 5$\sigma$ from -0.66 km s$^{-1}$ to 5.19 km s$^{-1}$. The color scale is stretched using the inverse hyperbolic sine function. The lowest contour emission corresponds to 5$\sigma$ (1$\sigma$ corresponds to 1.83 mJy beam$^{-1}$), with increments of 10$\sigma$. The field of view of the map is $6\arcsec\times6\arcsec$.} \label{fig:channel_map_c18o}
\end{figure*}

\section{JWST images}\label{JWST_image}
We compare the image of the $^{12}$CO molecular outflow with recent JWST images obtained with NIRCAM (PI K. Pontoppidan) centered over GSS30 IRS3.
\begin{figure*}[ht]
\includegraphics[width=0.45\textwidth]{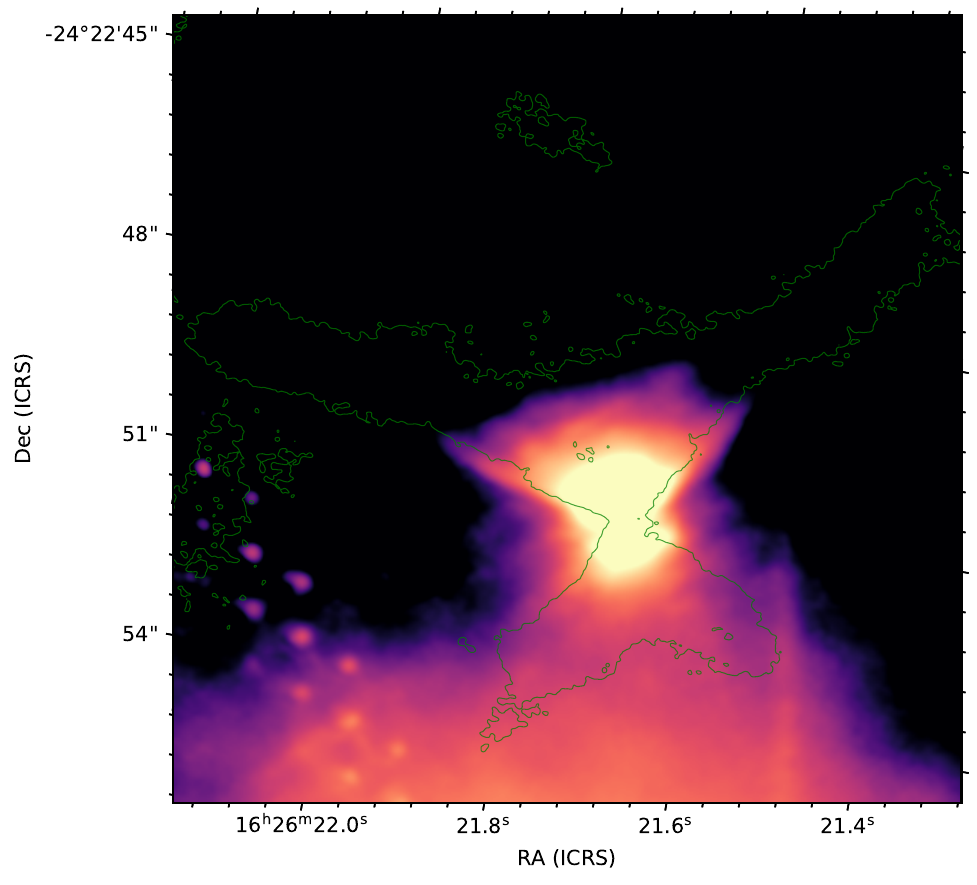}
\includegraphics[width=0.45\textwidth]{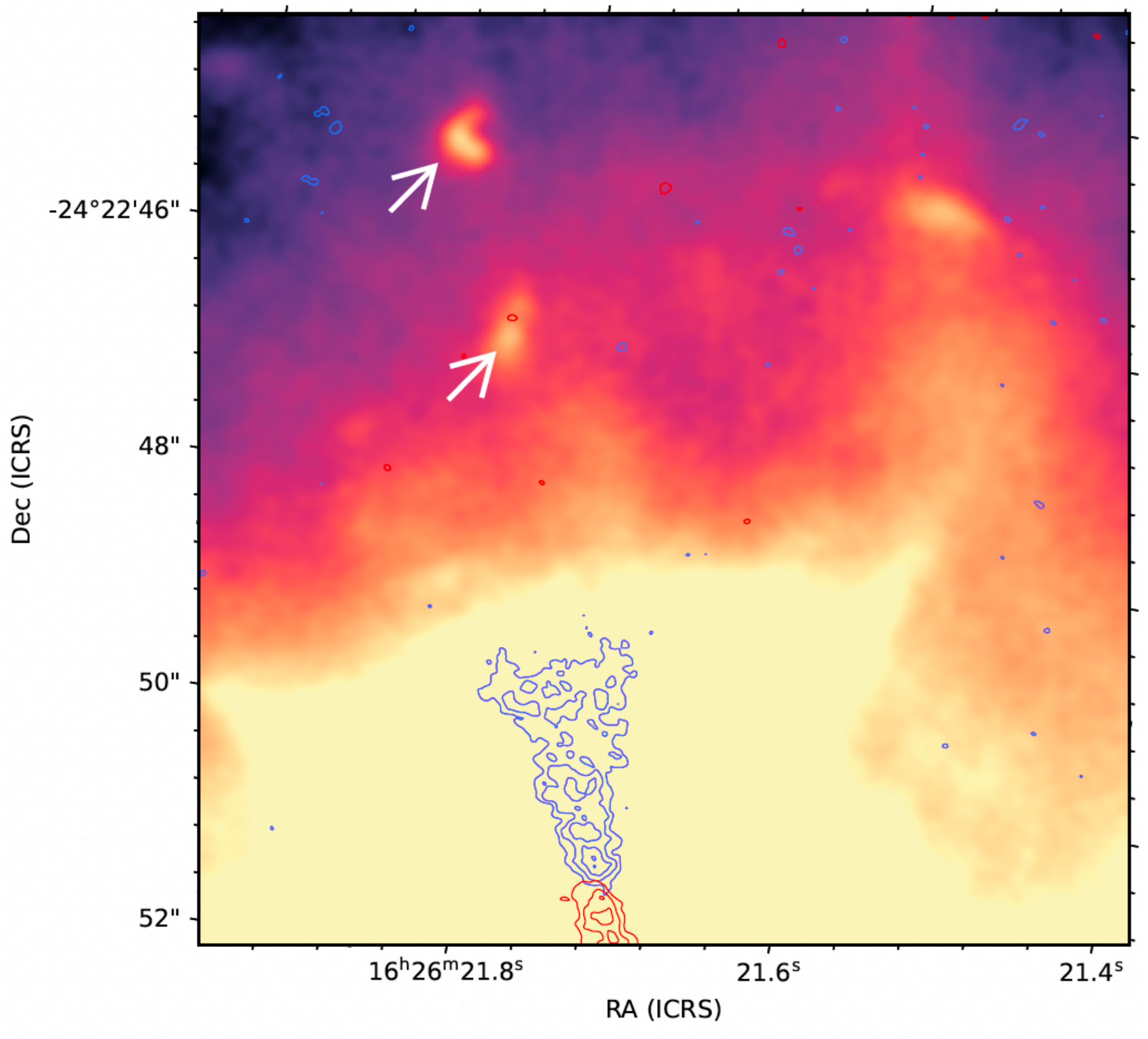} \\
\caption{JWST molecular hydrogen color image (4.7 $\mu$m image with the F470N filter) over GSS30 IRS3 ($^{12}$CO). Left panel shows the blueshifted component from the bottom-left panel of Figure \ref{fig:wide_collimated} with contours over 3 times the rms. Right panel only shows the collimated blueshifted and redshifted molecular outflow. Color image contrast limits are adjusted to enhance the presence of two knots in the direction of the blueshifted collimated molecular outflow emission which are pointed with two white arrows. The knots are located at 4$\farcs$9 and 6$\farcs$7 from the center of the continuum emission. Contours represent 3, 5, 7, 9, 11, 13 and 15 times the rms.}\label{fig:JWST_image}
\end{figure*}

\section{Other sources detected in the field of view}
\subsection{GSS30 IRS 1}\label{iras1}
\begin{figure}
\includegraphics[width=0.5\textwidth]{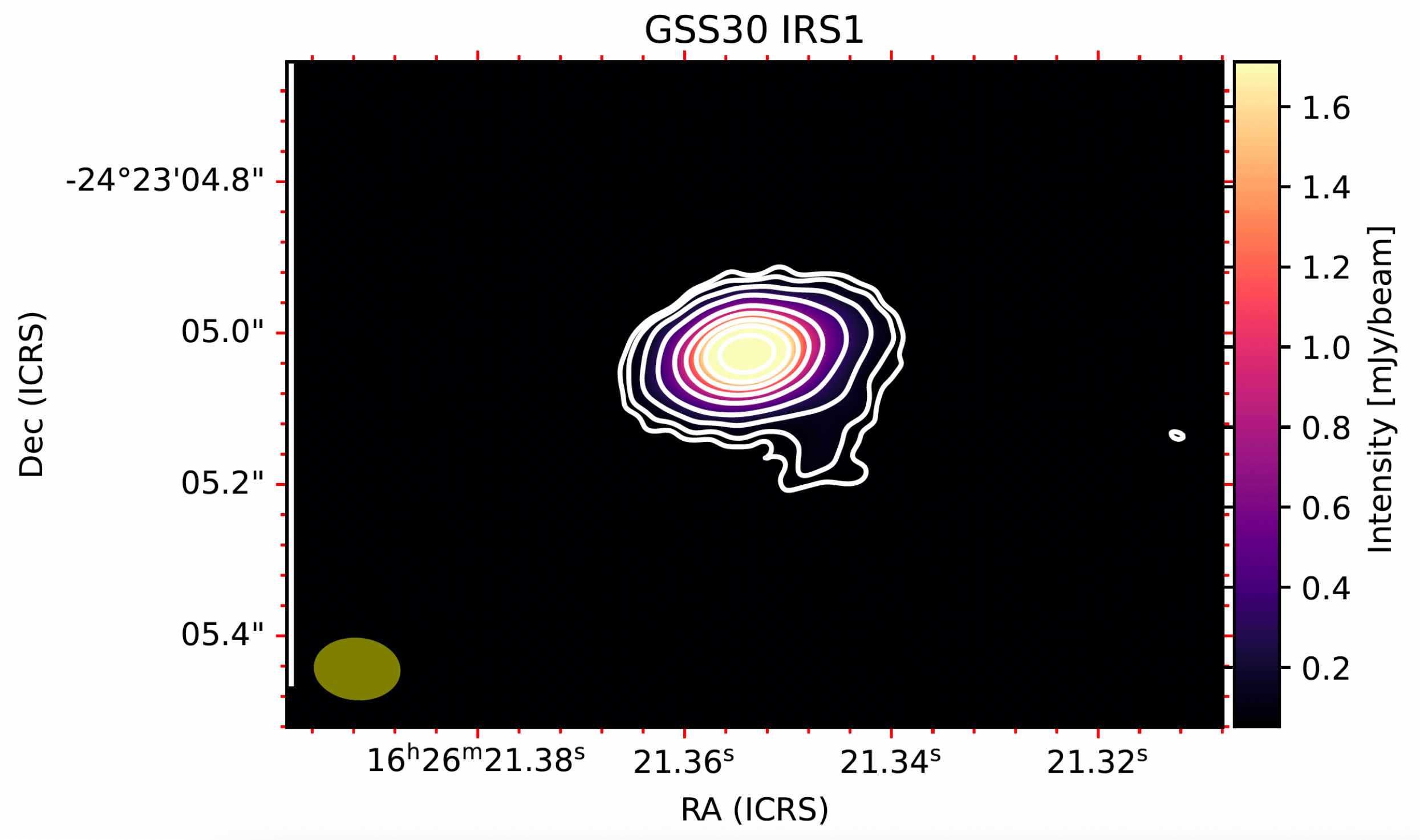}

\caption{ALMA continuum image of GSS30 IRS1 at 1.3 mm. White contours are 5, 10, 20, 40, 60, 80, 100, and 130 times the rms (1$\sigma$ = 17.1 $\mu$Jy beam$^{-1}$).}\label{irs1}
\end{figure}

We report the continuum detection of GSS30 IRS1, a Class I young stellar object \citep{Arnaud23}, located at a distance of approximately 14.7 arcseconds from GSS30 IRS3. Figure \ref{irs1} shows the 1.3 mm dust emission from GSS30 IRS1 with an angular resolution of 0$\farcs$052$\times$0$\farcs$042 (position angle of 2.5$\mathrm{^{o}}$), obtained using a robust parameter equal to 0.5. The continuum emission traces an inclined disk-like structure around the central protostar, with a small appendix similar to a spiral-like structure located at the west-south of the disk. The continuum emission measured above a 5$\sigma$ contour for the disk-like structure (excluding the spiral-like structure) has major and minor axes of 0\farcs34$\times$0\farcs22 ($\sim$67 $\times \sim$44 au), with a peak intensity and flux density of 2.72 mJy beam $^{-1}$ and 3.92 mJy, respectively. Coordinate of the peak intensity is R.A.=16h26m21.354s, Dec= -24$\mathrm{^{o}}$ 23'05$\farcs$03. The rms is 17.1 µJy beam s$^{-1}$. The inclination derived from the disk-like structure is approximately 50$\mathrm{^{o}}$.

Several emission lines, including $^{12}$CO (2-1), $^{13}$CO (2-1), C$^{18}$O (2-1), SO (6$_{5}$-5$_{4}$), and SiO (5-4), were detected surrounding GSS30 IRS1. A detailed analysis of the gas source and the spiral-like structure will be presented in a follow-up paper (Santamar\'ia-Miranda et al. in prep).

\subsection{GSS30 IRS2}\label{iras2}
\begin{figure*}
\includegraphics[width=0.5\textwidth]{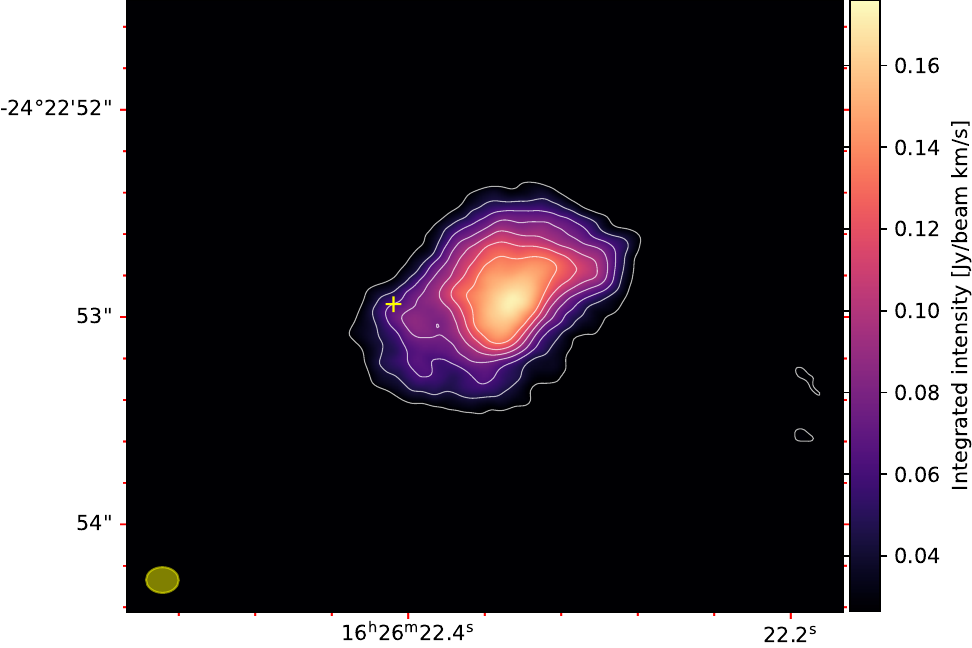} 
\includegraphics[width=0.5\textwidth]{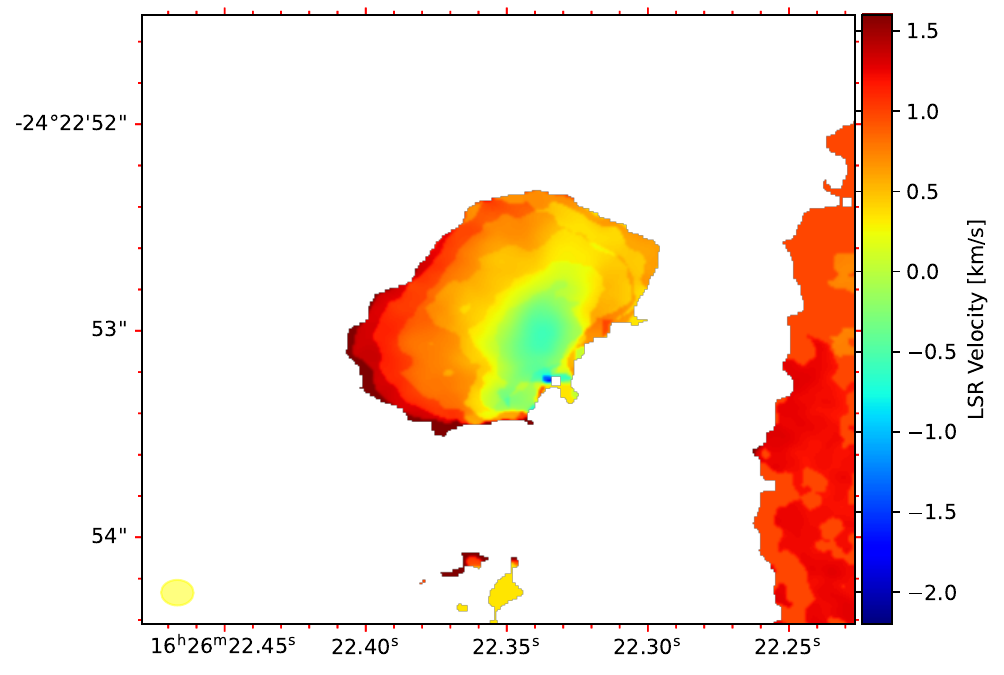} 

\caption{Left panel: $^{12}$CO emission line flux-integrated map of GSS30 IRS2. White contours represent the 3, 5, 7, 9, 11, 13, 15 times the rms that is 8.8 mJy beam$^{-1}$ km s$^{-1}$. Yellow cross represents the position of the central source. Right panel: $^{12}$CO (2-1) emission line intensity-weighted velocity map of GSS30 IRS2.}\label{irs2}
\end{figure*}

We report the detection of gas material surrounding the Weak-line T-Tauri star \citep{dolidze1959} GSS 30 IRS2 that is classified as M2 spectral type \citep{Greene95}. The $^{12}$CO emission is detected between -2.2 km/s and 1.6 km/s with respect to the local standard of velocity rest, and it integrated map is shown in Figure \ref{irs2}. Redshifted emission is very contaminated by cloud emission, and it is impossible to distinguish between emission from the source and from the cloud. The emission size is 1\farcs3$\times$0\farcs8 based on the 5$\sigma$ contour. There is not also very clear molecular outflow emission. Furthermore, there is no clear rotation in the intensity-weighted velocity map (Right panel in Figure \ref{irs2}). There is no detection of the continuum at 3$\sigma$ level (5.2 $\mu$Jy/beam) or any other of the emission lines. A summary of the main properties is shown in Table \ref{table2_iras2}.

\begin{table*}[h!] 
\footnotesize
\caption{GSS30 IRS2 main properties in the $^{12}$CO (2-1) emission line }
\label{table2_iras2}
\centering
\begin{tabular}{cccccccc}
 Frequency & Molecule & Transition & rms & Beam size & $\Delta$v &  Integrated intensity & Peak intensity \\

 [GHz]       &    & & [mJy beam$^{-1}$] & [$\arcsec$] & [km s$^{-1}$] & [Jy beam$^{-1}$ km s$^{-1}$] & [Jy beam$^{-1}$] \\

\hline \hline
230.53800 & $^{12}$CO & 2--1 & 1.0 & 0.158x0.124 & 3.8 & 3.51 & 0.172 \\ 
\end{tabular} 
\end{table*}

\section{SLAM fitting} \label{app:c18o_fitting}
\begin{figure*}[ht]
\includegraphics[width=0.45\textwidth]{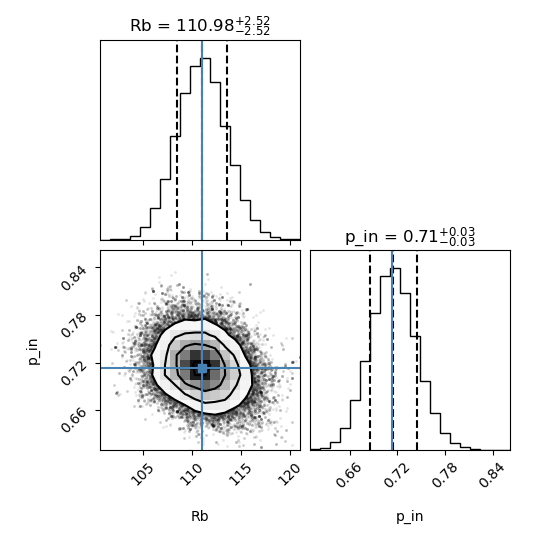}
\includegraphics[width=0.45\textwidth]{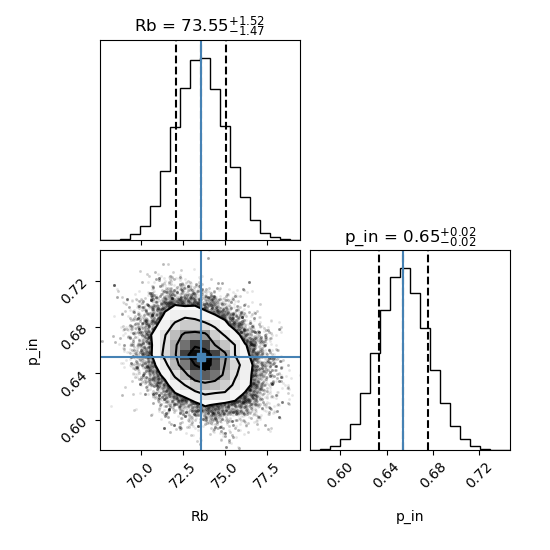} \\
\includegraphics[width=0.45\textwidth]{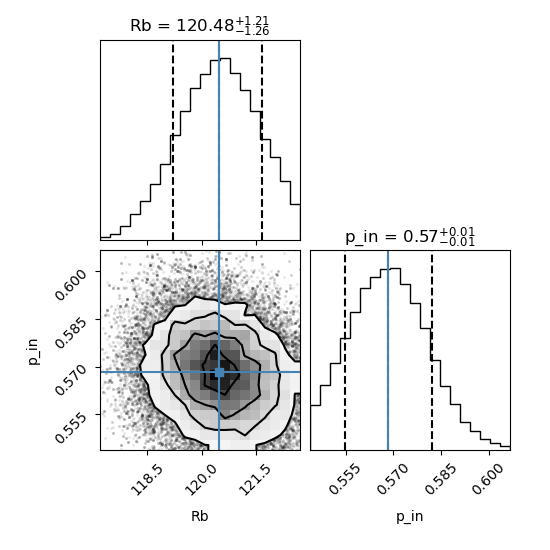}
\includegraphics[width=0.45\textwidth]{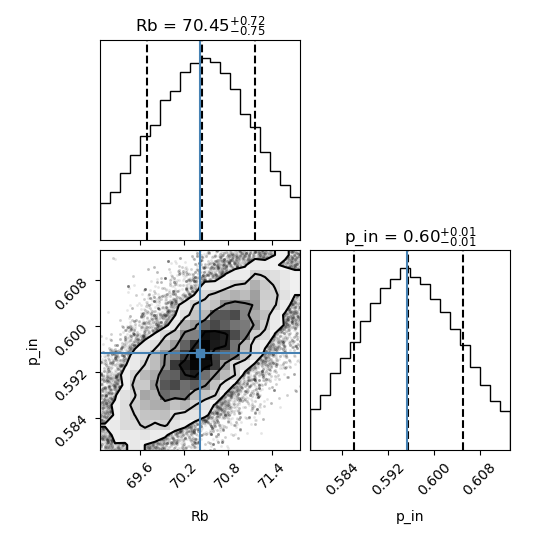} \\
\caption{Corner plot of the SLAM fitting using single-power law fitting. Top and bottom panels show the results for the C$^{18}$O (2-1) and $^{13}$CO (2-1) emission, and the left and right panels those with the edge and ridge methods, respectively. Dashed lines are the 16 and 84 percentiles. These plots shows that the prior ranges are wide enough to constrain the fitting parameters.}\label{corner_plot_SLAM}
\end{figure*}

\begin{figure}[ht]
\includegraphics[width=0.45\textwidth]{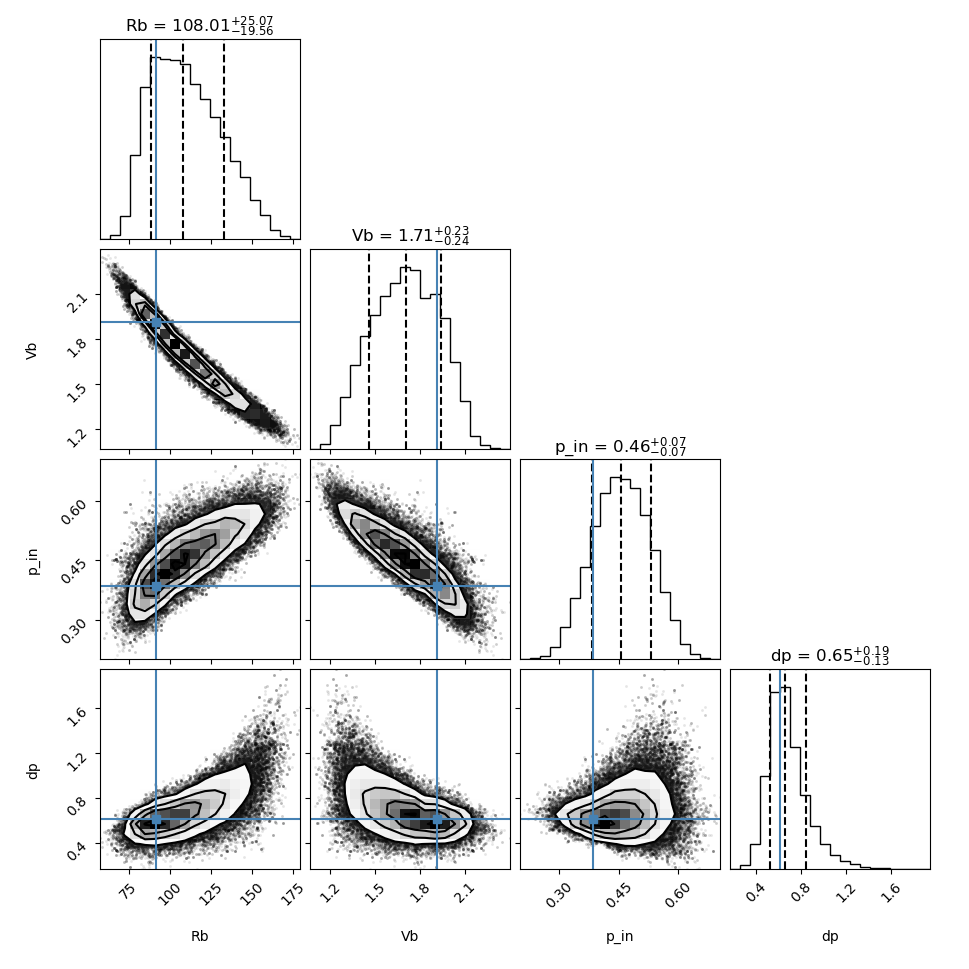} \\
\caption{Corner plot of the SLAM double-power law fitting for the C$^{18}$O (2-1) line using the edge method. Dashed lines are the 16 and 84 percentiles.}\label{corner_plot_SLAM_doble}
\end{figure}

This section shows the result of corner plot (Figures \ref{corner_plot_SLAM} and \ref{corner_plot_SLAM_doble}) of the \textit{SLAM} fitting in Section \ref{sec:SLAM_doble} for the C$^{18}$O and  $^{13}$CO lines.

\begin{figure*}[ht]
\includegraphics[width=0.45\textwidth]{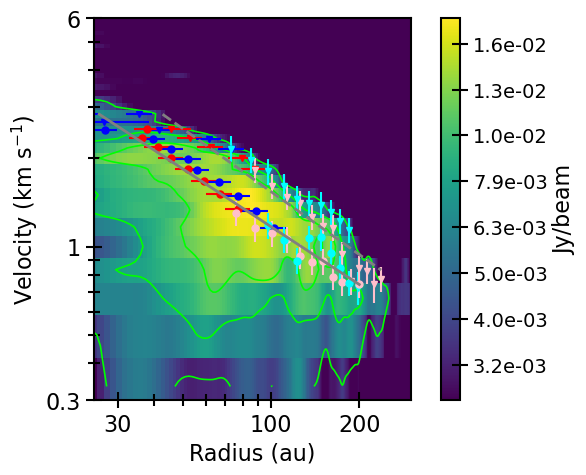}
\includegraphics[width=0.45\textwidth]{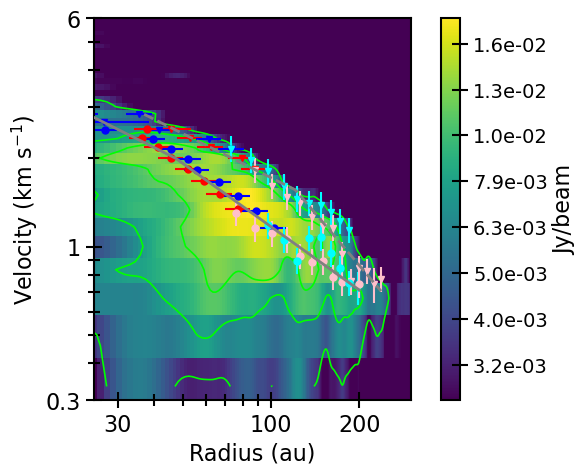} \\
\includegraphics[width=0.45\textwidth]{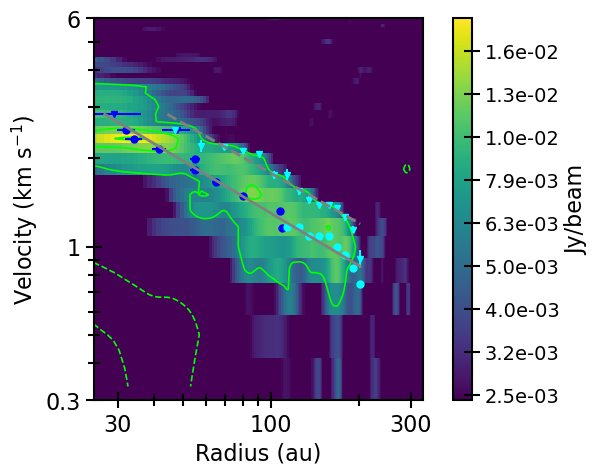}
\includegraphics[width=0.45\textwidth]{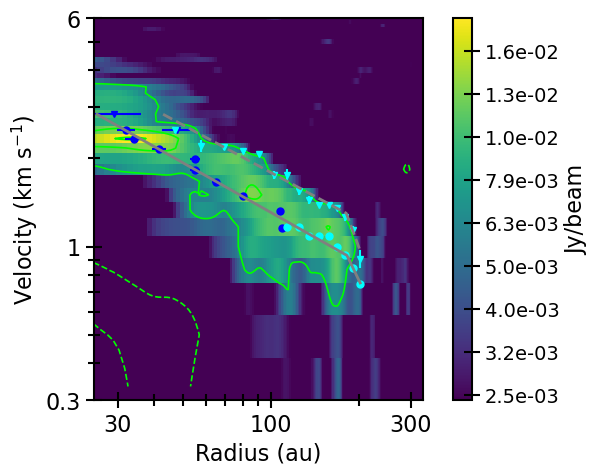} \\

\caption{Position velocity diagrams of the SLAM fitting in logarithm scale. Top and bottom panels show the results for the C$^{18}$O (2-1) and $^{13}$CO (2-1) emission, and the left and right panels those with single- and double-power law fitting, respectively. Symbols and colors are as shown in Figure \ref{fig:c18o_fitting_single}.}

\end{figure*}

\section{Dynamical parameters of the molecular outflow components} \label{dynamical_params}

In this subsection of the Appendix, we will analyze the dynamic properties of the three different components of the molecular outflow, separated in blueshited and redshifted emission with respect to the systemic velocity (2.84 km s$^{-1}$, as inferred in Section \ref{sec:SLAM_doble}). We will consider the emission inside the 3$\sigma$ contour level in Figure \ref{fig:wide_collimated} to perform the calculation in this subsection.  For inferring the dynamical properties of the molecular outflow, we will consider the $^{12}$CO (2-1) molecular line  since this transition clearly traces the different components of the outflow with good S/N, and the $^{13}$CO (2-1) line traces only the outflowing material closer to the central protostar together with a rotating disk.  
We calculate the dynamical time, the outflow mass, the mass-loss rate,  the outflow momentum, the kinetic energy, the mechanical luminosity,  and the outflow force.  

The dynamical time of the various outflow components was determined by considering the deprojected maximum velocity, the  deprojected length, and an outflow axis perpendicular to the dust disk, whose inclination is derived in Section \ref{sec:dust_emission}.

The maximum inferred dynamical time is approximately 535 years for the wider angle lower velocity blueshifted component. When considering both the blueshifted and redshifted emission, the average dynamical time for this component is around 340 years. We used the maximum velocity to calculate the dynamical time of each of the components. Then we averaged the dynamical time between the blue and red component. The intermediate velocity hourglass shape component exhibits an average dynamical time of approximately 43 years when considering both blueshifted and redshifted emission. Similarly, the most collimated inner high-velocity component shows an average dynamical time of approximately 31 years. 

The $^{12}$CO (2-1) emission line is optically thick at systemic velocities, since the brightness temperature of the line peak 57.9 K is similar to the kinetic temperature that we measured (58.7 K) from the linewidth. Therefore we determine the excitation temperature for the $^{12}$CO (2-1) line using the peak intensity obtained close to the position of the star as \begin{equation} T_{\mathrm{ex}} = \frac{h\nu_{12}/k}{\mathrm{ln}[1+ \frac{h\nu_{12}/k}{T_{o12}+J_{12}(2.7 K)}]}, \end{equation} where T$_{0}$ is the peak brightness temperature obtained close to the central continuum peak position. We derived a T$_{\mathrm{ex}}$ for the $^{12}$CO (2-1) of 25.7 K, and we assumed this value for all the molecular outflow components. 
 
Then, we followed the prescription in \citet{Scoville1986} and \citet{Palau2007} to calculate the column density and the mass of the outflow. We derived the mean optical depth in the wings of the $^{12}$CO (2-1) emission line. We define the line wings as [-19.9 km/s, -18,7 km/s] and [23.2 km/s, 23.8 km/s] for the inner molecular outflow, [-18.0 km/s, -17.4 km/s] and [21.9 km/s, 22.6 km/s] for the intermediate hourglass shape and only two channels (-1.6 km/s and 7.9 km/s)  for the wider angle low velocity molecular outflow. The optical depth is obtained for each of the three molecular outflow components, obtaining  values between 0.04 and 0.51, in the optically thin regime. Then we assume that the emission is optically thin in the line wings. We infer a column density $\sim$2$\times$10$^{16}$cm$^{2}$ for the wider angle and the inner molecular outflow components and a value one order of magnitude smaller for the highly collimated high-velocity molecular outflow component. Finally, we calculated the outflow mass using the inferred column density, the emission area, and a CO abundance of 10$^{-4}$.

We derived the rest of the parameters following the formulas in \citet[Table 3]{Santamaria-Miranda20-1} correcting for inclination. The total outflow mass of the molecular outflow is 6.1$\times10^{-5}$ M$_{\odot}$, the momentum varies between 9.8$\times10^{-6}$  M$_{\odot}$ km s$^{-1}$ and 4.5$\times10^{-3}$ M$_{\odot}$ km s$^{-1}$, the kinetic energy varies between 4.7 $\times10^{39}$ erg and  4.7 $\times10^{41}$ erg, and the mechanical force varies between 4.8$\times10^{-7}$ M$_{\odot}$ km s$^{-1}$ yr$^{-1}$ and 1.0$\times10^{-5}$ M$_{\odot}$ km s$^{-1}$ yr$^{-1}$. All these parameters are summarized in Table  \ref{tab:dynamical_parameters}.

These results are affected by several assumptions that may affect the estimated mass. Previous studies point out that the optical depth of molecular outflows is underestimated, and masses should be at least a factor two higher \citep{Dunham2014} or five \citep{marel13}. The first source of uncertainty is T$_{ex}$; our value is in the expected range for molecular outflows (T$_{\mathrm{ex}}$ = 10-50 K \citealt{Bally2016}); if we use a T$_{ex}$=50 K, the mass will increase by a factor of 1.4 with respect to that with T$_{ex}$=25.7 K. Another source of uncertainties is related to the parental cloud material that contaminates the emission of the molecular outflow close to the systemic velocity, needing the exclusion of those channels for the dynamical parameters calculations.  \citet{Offner-2011}, using synthetic observations, found that the missed outflow mass might increase by a factor 5 to 10 within 2 km s$^{-1}$ from the rest velocity. However, the most significant uncertainty factor in our calculations is the interferometric filtering. It is noticeable when comparing our dynamical properties with the ones obtained in low-mass Class 0 molecular outflows observed with single-dish telescopes. For example, the outflow mass in Class 0 and I sources is in the range between $\times10^{-3}$ M$_{\odot}$ and $\times10^{-1}$ M$_{\odot}$  \citep{Mattrom2017}, and our total outflow mass is below these values. Another example is the mechanical force  expected to be between  1.0$\times10^{-5}$ M$_{\odot}$ year$^{-1}$ km  s$^{-1}$ and $\times10^{-2}$ M$_{\odot}$ yr$^{-1}$ km s$^{-1}$ for Class 0 sources \citep{Mattrom2017} and the mechanical force of GSS30 IRS3 outflow is in the lower limit.
Therefore, we conclude that all derived parameters may be treated as lower limits.

\section{Radial brightness profile fitting}
\label{Analysis:asymmetries}
To further investigate if the bumps detected in the major axis of the radial intensity profile of the dust disk surrounding GSS30 IRS3 are real features and not a byproduct of the cleaning process, we fit a 1D profile to the ALMA visibilities with a nonparametric model using the \textit{Frank} library \citep{Jennings2020}, which provides a super-resolution brightness profile reconstruction. A major advantage of fitting visibilities is avoiding false structures created during the CLEAN process. Data were averaged in frequency to get spectral windows of 8 channels to reduce data volume and speed up calculations, excluding baseband 1 and 4, which are dedicated to detecting molecular emission lines or has a strong detected line where the contribution to the continuum was negligible. \textit{Frank} can estimate the inclination and position angle of the protoplanetary disk; however, \citet{Benisty2021} noted that \textit{Frank} does not fit these two parameters accurately, so we fixed them to an inclination of 64.30 deg and a position angle of 109 deg, as inferred in Section \ref{sec:dust_emission} considering the emission above the 5$\sigma$ contour.

\begin{figure*}
\includegraphics[height=0.3\textwidth]{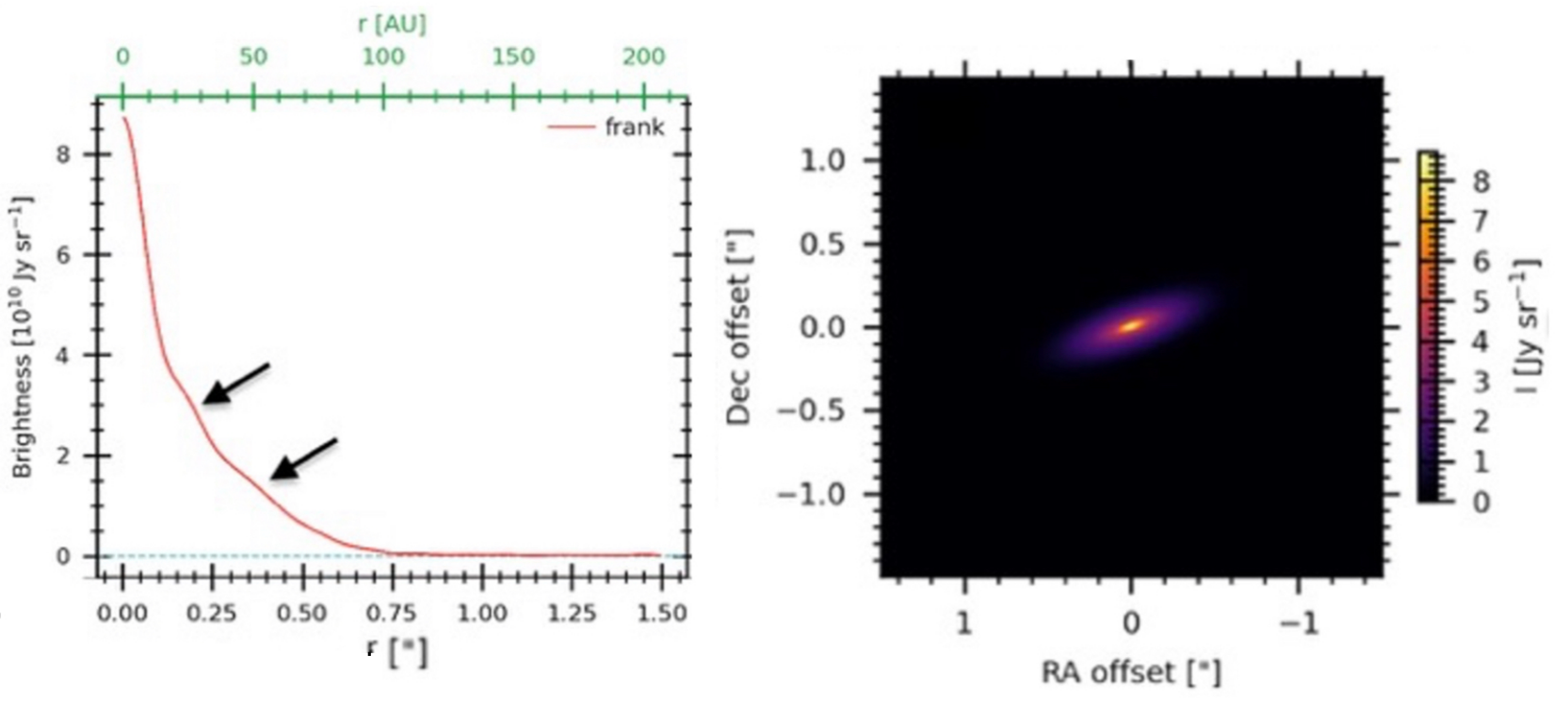}
\caption{Results of the best one-dimension \textit{Frank} fit.  The left panel is the fitted radial intensity frank profile. The position of the bumps is marked with two black arrows. The right panel is the non-convolved Frank profile reprojected over two dimensions.} \label{frank_fit_v2}
\end{figure*}

Before fitting a Bessel functions to the brightness profile, \textit{Frank} applies a power spectrum estimate as a prior. There are several hyperparameters that affect this power spectrum, but the two most relevant ones are $\alpha$ and w$\mathrm{_{smooth}}$.  $\alpha$  works as an S/N threshold for the maximum baseline where visibilities are fitted (a higher value implies a stricter S/N threshold), while w$\mathrm{_{smooth}}$ affects the smoothness of the power spectrum, where a higher value implies a smoother power spectrum. We explore the fitting quality using different hyperparameters values ($\alpha$ between 1.05 and 1.30 and w$\mathrm{_{smooth}}$ between 10$^{-4}$ and 10$^{-1}$), not finding any substantial difference between the different fittings. We fixed the hyperparameters to $\alpha$= 1.30 and w$\mathrm{_{smooth}}$=10$^{-4}$ because they present the least noisy results. The best-fit model (Figure \ref{frank_fit_v2}) showed the presence of two bumps. 
To obtain the bumps size and location, we fit an exponential to the brightness profile obtained from \textit{Frank} to isolate the bumps. Then, we fitted Gaussians around the areas where the bumps are located. Their positions are $\sim$0\farcs19 ($\sim$26 au) and  $\sim$0\farcs36 ($\sim$50 au) from the center of the disk, 0\farcs09  ($\sim$13.6 au) and 0\farcs14 ($\sim$19 au) wide, respectively.

\end{appendix}

\end{document}